\documentclass[fleqn,10pt]{wlscirep}
\usepackage[utf8]{inputenc}
\usepackage[T1]{fontenc}

\usepackage{amssymb}
\usepackage{amsfonts}
\usepackage{bm}

\usepackage{xr}
\newcommand{\dd}{\text{ d}}

\usepackage[nomarkers,figuresonly,nolists]{endfloat}
\usepackage{xcolor}

\title{Prediction of Optimal Drug Schedules for Controlling Autophagy}

\author[1,+]{Afroza Shirin}
\author[1,+]{Isaac S. Klickstein}
\author[2,+]{Song Feng}
\author[2,+]{Yen Ting Lin}
\author[2,+,*]{William S. Hlavacek}
\author[1,+,*]{Francesco Sorrentino}

\affil[1]{Mechanical Engineering Department, University of New Mexico, Albuquerque, NM 87131}
\affil[2]{Theoretical Biology and Biophysics Group, Theoretical Division and Center for Nonlinear Studies, Los Alamos National Laboratory, Los Alamos, NM 87545}
\affil[+]{These authors contributed equally.}
\affil[*]{Corresponding authors: wish@lanl.gov (W.S.H.), fsorrent@unm.edu (F.S.)}

\begin{abstract}
The effects of molecularly targeted drug perturbations on cellular activities and fates are difficult to predict using intuition alone because of the complex behaviors of cellular regulatory networks. An approach to overcoming this problem is to develop mathematical models for predicting drug effects. Such an approach beckons for co-development of computational methods for extracting insights useful for guiding therapy selection and optimizing drug scheduling. Here, we present and evaluate a generalizable strategy for identifying drug dosing schedules that minimize the amount of drug needed to achieve sustained suppression or elevation of an important cellular activity/process, the recycling of cytoplasmic contents through (macro)autophagy. Therapeutic targeting of autophagy is currently being evaluated in diverse clinical trials but without the benefit of a control engineering perspective. Using a nonlinear ordinary differential equation (ODE) model that accounts for activating and inhibiting influences among protein and lipid kinases that regulate autophagy (MTORC1, ULK1, AMPK and VPS34) and methods guaranteed to find locally optimal control strategies, we find optimal drug dosing schedules (open-loop controllers) for each of six classes of drugs and drug pairs. Our approach is generalizable to designing monotherapy and multi therapy drug schedules that affect different cell signaling networks of interest. 
\end{abstract}

\begin{document}

\flushbottom
\maketitle
%
%
\thispagestyle{empty}


\section*{Introduction}
Although there is much current interest in using combinations of molecularly targeted drugs to improve outcomes for cancer patients \cite{Jameson2015,Gatzka2018}, relatively little work has been done in the area of formal therapy design, meaning therapy selection and/or scheduling driven by insights from mathematical models \cite{Anderson2008,Michor2015}. Formal approaches to therapy design are potentially useful for at least three reasons. First, all possible combinations of drugs may be difficult, if not impossible, to evaluate experimentally simply because of the large number of possible combinations. Second, an ability to extrapolate accurately beyond well-characterized scenarios with the aid of predictive models would be valuable for individualized treatment, especially in cases where molecular causes of disease are diverse and vary from patient to patient, as in many forms of cancer \cite{Vogelstein2013}. Third, it is often non-obvious how the immediate effects of drug perturbations propagate through a cellular regulatory network to affect cellular phenotypes and fates \cite{Rukhlenko2018} or how drug combinations might be deployed to avoid or delay the emergence of resistance, a common response of malignant cells to targeted therapies \cite{Ramos2015}. Predictive models promise to help identify new robust therapies.

Here, we apply mathematical modeling and optimal control methods to design drug schedules for manipulating autophagy, a stress-relieving/homeostatic cellular recycling process that, when nutrients are in limited supply, generates building blocks for protein synthesis through degradation of cytoplasmic contents \cite{Klionsky2000}, such as cytotoxic protein aggregates that are too large for proteosomal degradation and damaged organelles (e.g., depolarized mitochondria). Autophagy also plays an important role in immunity \cite{Deretic2013,Levine2011}; the autophagic degradative machinery can be directed to target intracellular microbes, such as {\it Mycobacterium tuberculosis}, for destruction. 

Cytoplasmic contents that are targeted for autophagic degradation are first trapped in double-membrane vesicles, termed autophagosomes or autophagic vesicles (AVs), and then delivered to lysosomes for digestion \cite{mizushima2008autophagy,nakatogawa2009dynamics}. The production of AVs is controlled by an intricate regulatory network, in which three protein kinase-containing complexes are prominent: the heterotrimeric AMP-activated kinase (AMPK), which senses energy (glucose) supply through interactions with adenosine derivatives (AMP and ATP) \cite{kahn2005amp,loffler2011ulk1}; MTOR complex 1 (MTORC1), which senses amino acid supply and growth factor signaling through interactions with small GTPases localized to lysosomal surfaces (Rag proteins and RHEB) \cite{zoncu2011mtor,nazio2013mtor}; and the ULK1 complex, which is activated by AMPK and repressed by MTORC1 \cite{shang2011nutrient,dunlop2011ulk1,kim2011ampk}. A fourth complex, which contains a lipid kinase, VPS34, also plays an important role \cite{fimia2007ambra1,di2010dynamic}. Interestingly, VPS34 and MTOR are phylogenetically related: they are both members of the phosphoinositide 3-kinase (PI3K) family. Drugs with specificity for each of these kinases are available, and because of the relationship between MTOR and VPS34, drugs are also available with dual specifity for this pair of kinases \cite{Galluzzi2017,Moschetta2014,Hardie2013}.

In cancer, and other contexts, autophagy is a double-edged sword \cite{Shintani2004}. It can protect cancer cells from stresses of the tumor environment (e.g., lack of nutrients because of defective vasculature) or induce cell death if recycling is excessive. Thus, there are potential benefits to be gained by using drugs to either upregulate autophagy (to kill malignant cells through excessive recycling) or downregulate autophagy (to kill cancer cells that rely on autophagy for survival) \cite{MulcahyLevy2017}. 

To investigate how single drugs and drug pairs might be best used for these purposes, we constructed a system of nonlinear ordinary differential equations (ODE) that captures regulatory interactions between MTORC1, ULK1, AMPK, and VPS34, as well as the idealized pharmacokinetics of kinase inhibitors specific for MTORC1, ULK1, AMPK, and VPS34, such as rapamycin \cite{edwards2007rapamycin}, SBI-0206965 \cite{egan2015small}, dorsomorphin \cite{meley2006amp}, and SAR405 \cite{ronan2014highly}, respectively. We also considered an allosteric activator of AMPK (e.g., PF-06409577\cite{cameron2016discovery}) and a kinase inhibitor with dual specificity for MTORC1 and VPS34 (e.g., buparlisib\cite{burger2011identification}). Although the model is minimalist by design, it reproduces key behavioral features of earlier, more mechanistically detailed models \cite{szymanska2015computational,martin2013computational}, such as oscillatory responses to intermediate levels of nutrient or energy stress. We then applied optimization methods implemented in the open-source $\mathcal{PSOPT}$ software package \cite{becerra2010solving} to find locally optimal dosing schedules that minimize the total amount of drug needed to drive the network to a desired, non-attracting operating point (corresponding to low or high AV count/turnover) and maintain it there. The dosing schedules are non-obvious, and synergistic drug pairs were predicted (drug 6 plus drug 1, 2 or 3), such as the combination of a VPS34 inhibitor and a dual specificity PI3K inhibitor, which acts on both VPS34 and MTORC1. This drug pair requires less total drug to achieve the same effect than either of the individual drugs alone and is relatively fast acting, which may be important for preventing or slowing the emergence of resistance. 

The approach illustrated here differs from earlier applications of control theory concepts in the area of formal therapy design \cite{martin1992optimal,swierniak2003optimal,ledzewicz2006drug,ledzewicz2007optimal,joshi2002optimal} in that 1) the system being controlled is a cellular regulatory network, 2) the control interventions are injections (i.e., inputs) of (combinations of) molecularly targeted drugs, and 3) the control objective is manipulation of a cellular phenotype, namely the number of AVs per cell, which is related to the rate of AV turnover, with minimization of total drug used and a constraint on the maximum instantaneous drug concentration. The rationale for minimizing drug use is to avoid offtarget effects and associated toxicities. Our work is distinct from earlier studies of (non-biological) nonlinear network control \cite{cornelius2013realistic,wang2016geometrical,zanudo2017structure,klickstein2017locally}, in that our control goal is not to drive the system to an attractor (e.g., a stable steady state or limit cycle), but to an arbitrary point in phase space (i.e., the multidimensional space defined by the state variables of a system) and to then maintain the system there indefinitely. The approach is both flexible and generalizable and provides a means for computationally prioritizing drug dosing schedules for experimental evaluation. 
\section*{Results}
\subsection*{Model for cellular regulation of autophagy and the effects of targeted drug interventions}\label{model}

\begin{figure}[t!]
	\centering
	\includegraphics[width=\linewidth]{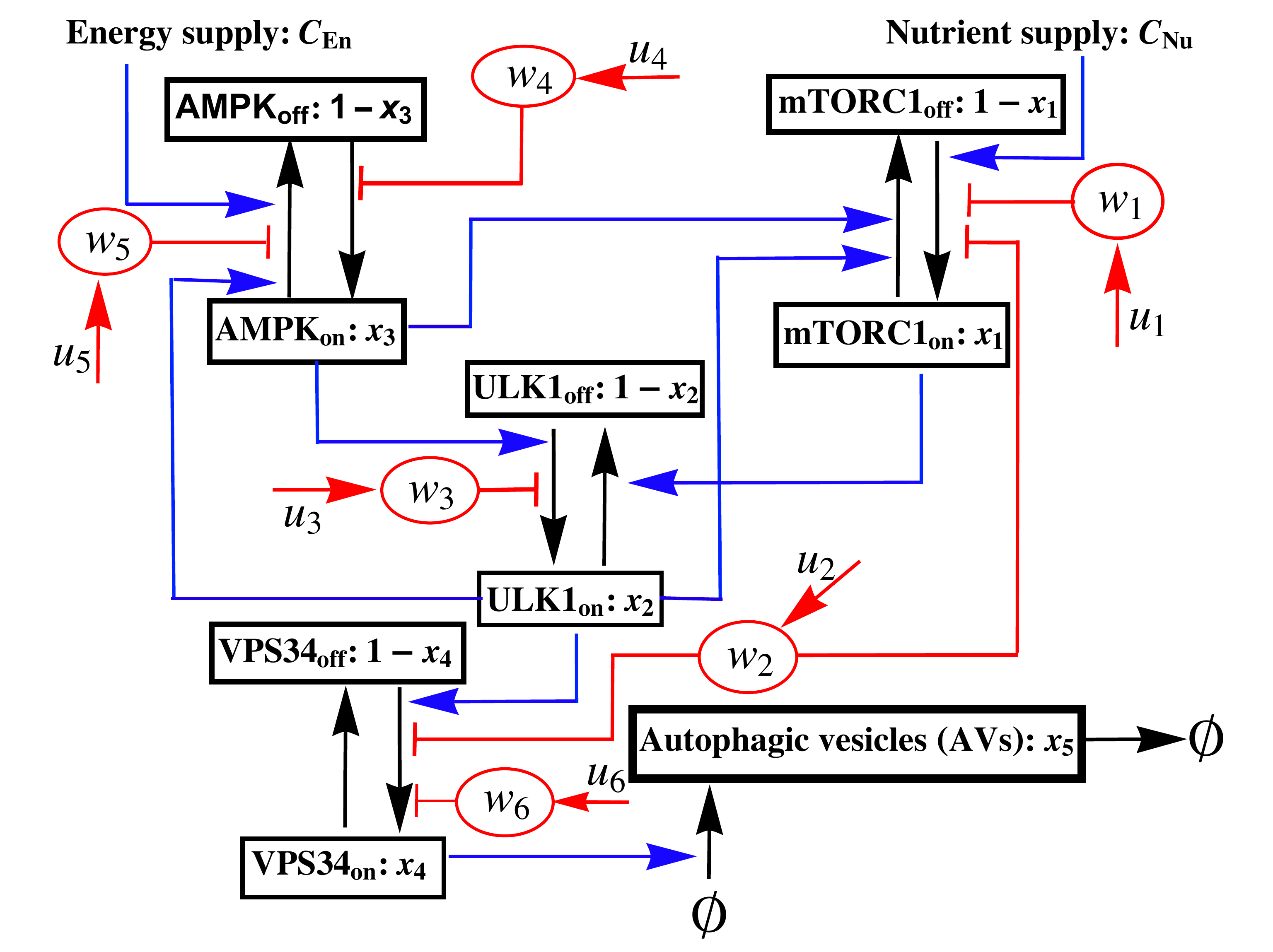}
	\caption{Schematic diagram of a minimalist mathematical model for regulation of autophagy and the effects of targeted drug interventions.
The model accounts for two physiological inputs (energy and nutrient supply) and regulatory influences, stimulatory or inhibitory, within a network of interacting kinases. Each kinase is taken to have a constant total abundance and to be dynamically distributed between active and inactive forms. The active fractions of MTORC1, ULK1, AMPK, and VPS34 are represented by $x_1$, $x_2$, $x_3$ and $x_4$, respectively. Targeted drugs, denoted by red ovals, promote kinase inactivation or activation as indicated. Six drug types are considered: 1) a kinase inhibitor specific for MTORC1, 2) a kinase inhibitor specific for both MTORC1 and VPS34, 3) an ULK1 kinase inhibitor, 4) an allosteric activator of AMPK, 5) an AMPK kinase inhibitor, and 6) a VPS34 kinase inhibitor. The supplies of cellular energy and nutrients ($C_{\rm En}$ and $C_{\rm Nu}$), together with drug concentrations ($w_1,\ldots,w_6$), determine the kinase activities of MTORC1, ULK1, AMPK, and VPS34 and thereby the rate of synthesis of autophagic vesicles (AVs). The control parameters are drug injection/input rates ($u_1,\ldots,u_6$). Note that drug clearance is not indicated in this diagram but is considered in the model equations.}
	\label{fig:model}
\end{figure}

A prerequisite for formal therapy design is a mathematical model that captures the relevant effects of drugs of interest. Given our interest in using drugs to modify the  process of (macro)autophagy, we constructed a model for regulation of the rate of synthesis of autophagic vesicles (AVs) that accounts for the enzymatic activities and interactions of four kinases that play critical roles in regulating autophagy, all of which are potential drug targets. The model further considers the effects of achievable drug interventions and idealized drug pharmacokinetics, meaning instantaneous drug injection according to a time-dependent control function and first-order clearance. The model is illustrated in Fig.~\ref{fig:model}.

The model was constructed in two steps. First, we constructed a minimalist model for physiological regulation of autophagy consistent with key features of earlier, more mechanistically detailed models \cite{martin2013computational,szymanska2015computational} (see ``Formulation of the Model'' in Supplementary Methods for details). These features include the time scale of drug-stimulated autophagy induction and the dynamic range of regulation characterized by Martin \textit{et al.}\cite{martin2013computational} and the qualitative system behaviors characterized by Szyma{\'n}ska \textit{et al.}\cite{szymanska2015computational}, including a steady, low level of autophagy at low stress levels, oscillatory behavior at intermediate stress levels, and a steady, high level of autophagy at high stress levels. Simulations based on the present model---generated through numerical integration of the equations given below---and simulations based on earlier, related models\cite{martin2013computational,szymanska2015computational} are compared in Supplementary Fig.\ S1. Simulations of AV dynamics are compared to measured AV dynamics\cite{szymanska2015computational} in Supplementary Fig.\ S2.

The model of Fig.~\ref{fig:model} is intended to provide an idealized representation of regulation of AV synthesis in a single (average) cell in response to changes in the cellular supplies of energy and nutrients, which are treated in the model as external inputs that modulate the serine/threonine-specific protein kinase activities of AMPK and MTORC1, respectively. Thus, the model reflects regulation of AMPK activity by the cellular AMP:ATP ratio, which is affected by glucose availability, for example, and regulation of MTORC1 activity via, for example, the various amino acid-sensing regulators of Ragulator-associated heterodimeric Rag proteins, which recruit MTORC1 to lysosomes for activation in a manner that depends on their regulated guanine nucleotide binding states. The model further accounts for regulatory interactions among AMPK, MTORC1, a third serine/threonine-specific protein kinase ULK1, and a class III phosphoinositide 3-kinase (PI3K) VPS34. As noted earlier, these kinases are key regulators of autophagy, and each is a potential drug target. 

In the second step of model construction, we added idealized consideration of six distinct drug interventions, which correspond to interventions achievable through use of available small-molecule compounds, such as rapamycin\cite{edwards2007rapamycin} (an inhibitor of MTORC1 kinase activity), buparlisib\cite{burger2011identification} (an inhibitor of PI3K-family kinases that has specificity for both MTORC1 and VPS34), SBI-206965\cite{egan2015small} (an inhibitor of ULK1 kinase activity), dorsomorphin\cite{meley2006amp} (an inhibitor of AMPK kinase activity), PF-06409577\cite{cameron2016discovery} (a direct activator of AMPK kinase activity), and SAR405\cite{ronan2014highly} (an inhibitor of VPS34 kinase activity). Each drug $i \in \{1,\ldots,6\}$ (Fig.~\ref{fig:model}) is taken to be cleared via a pseudo first-order process and introduced in accordance with a specified, time-dependent injection function $u_i$.

The model was formulated as a coupled system of nonlinear ordinary differential equations (ODEs):
\begin{subequations}{\label{eq:ode}}
	\begin{align}
	T \dot{x}_1(t) & = (1-x_1)C_\text{Nu} H(w_1) H(w_2) - x_1 h_{12}(x_2)h_{13}(x_3),\label{eq:drug1and2}\\
	T  \dot{x}_2(t) & = (1-x_2) h_{23}(x_3) H(w_3) - x_2 h_{21}(x_1),\\
	T  \dot{x}_3(t) & = (1-x_3) k_1 H(w_4) - C_\text{En}x_2 x_3 H(w_5),\\
	T \dot{x}_4(t) & = (1-x_4)h_{42}(x_2) H(w_2)H(w_6) - k_2x_4, \label{eq:drug2and6}\\
	T \dot{x}_5(t) & = k_3x_4 - k_4 x_5,\\
	T \dot{w}_i(t) & = b_i u_i(t) - \delta_i w_i(t),
	\ i=1,\ldots,6.\label{eq:bi}
	\end{align}
\end{subequations} 
In these equations, $t$ is time (in min) and $T$ is a timescale, which we specify as $1.0$ min. The variable $x_1$ represents the fraction of MTORC1 that is active, the variable $x_2$ represents the fraction of ULK1 that is active, the variable $x_3$ represents the fraction of AMPK that is active, the variable $x_4$ represents the fraction of VPS34 that is active, and the variable $x_5$ represents the AV count or number of AVs per cell (on a continuum scale). Thus, $x_i$ always lies somewhere in the interval $[0,1]$ for $i=1,\ldots,4$. 
The AV count is bounded $0 \leq x_5 \leq k_3/k_4$ because $x_5(t) = 0$ implies $\dot{x}_5(t) \geq 0$ and $x_5(t) = k_3/k_4$ implies $\dot{x}_5 \leq 0$ (by the previously stated bound on $x_4(t)$).  
The variables $w_1,\ldots,w_6$ represent the dimensionless concentrations of drugs 1--6. Thus, $w_i \geq 0$ for each $i$. The non-dimensional parameters $C_{\rm En}$ and $C_{\rm Nu}$ are condition-dependent constants that define the supplies of energy and nutrients. An increase in energy supply is taken to positively influence the rate of deactivation of AMPK, and an increase in nutrient supply is taken to positively influence the rate of activation of MTORC1. The non-dimensional parameters $k_1$ and $k_2$ influence the rate of activation of AMPK and the rate of deactivation of VPS34, respectively. The non-dimensional parameter $k_3$ is the maximal rate of VPS34-dependent synthesis of AVs, and the non-dimensional parameter $k_4$ is the rate constant for clearance of AVs. Taking the rate of AV synthesis to be proportional to VPS34 activity is consistent with the model of Martin \textit{et al.}\cite{martin2013computational}, as is (pseudo) first-order clearance of AVs. The non-dimensional parameters $\delta_1,\ldots,\delta_6$ are rate constants for clearance of drugs 1--6.  Each $h_{ji}(x_i)$ is a non-dimensional Hill function that has the following form:
\begin{equation} \label{eq:Hill_h}
h_{ji}(x_i) = r_{b,ji} + \left(r_{m,ji}-r_{b,ji}\right)\frac{ x_i^{n_{ji}}}{x_i^{n_{ji}}+\theta_{ji}^{n_{ji}}}
\end{equation}
where $n_{ji}$ (the Hill coefficient), $r_{b,ji}$, $r_{m,ji}$ and $\theta_{ji}$ are non-negative constants. The $h$ functions account for regulatory influences among the four kinases considered in the model; the influences considered are the same as those considered in the model of Szyma{\'n}ska \textit{et al.}\cite{szymanska2015computational} (cf. Fig.~\ref{fig:model} and Figs. 1 and 2 in Ref. 33). Each $H(w_i)$ is a non-dimensional Hill function that has the following form:
\begin{equation} \label{eq:Hill_H}
H(w_i) = r_{m} -  \left(r_{m}-r_{b} \right) \frac{w_i^{n} }{w_i^{n}+\theta^{n}}
\end{equation}
where $n$ (the Hill coefficient), $r_{b}$, $r_{m}$ and $\theta$ are non-negative constants. The $H$ functions account for drug effects on kinase activities. The parameters $b_i$ ($i=1,\ldots,6$) in Eq.\ \eqref{eq:bi} are Boolean variables introduced for convenience, for the purpose of defining allowable drug combinations. Recall that the $u_i$ terms represent drug injection/input functions, which will be determined by solving an optimal control problem (described in the following section). 

Parameter settings are summarized in Supplementary Tables S1 and S2. Each $\delta$ parameter was assigned a value consistent with a known drug half-life\cite{cameron2016discovery,sato2006temporal,baselga2017buparlisib,milkiewicz2011improvement,engers2013synthesis,juric2017first} (Supplementary Table S2). Other parameters were assigned values that allow the model to reproduce the qualitative signaling behaviors of the AMPK-MTORC1-ULK1 triad characterized in the theoretical study of Szyma{\'{n}}ska \textit{et al.}\cite{szymanska2015computational} and to reproduce the timescale of autophagy induction and the range of regulation quantified experimentally in the study of Martin \textit{et al.}\cite{martin2013computational}. According to Szyma{\'{n}}ska \textit{et al.}\cite{szymanska2015computational}, at low levels of energy/nutrient stress, ULK1 activity, which can be expected to correlate with autophagic flux and AV count, is steady and low; at intermediate levels of stress, ULK1 activity is oscillatory; and at high levels of stress, ULK1 activity is steady and high. As noted earlier, in Supplementary Fig.\ S1, we compare simulations based on Eq.\ \eqref{eq:ode} with simulations based on models of Szyma{\'{n}}ska \textit{et al.}\cite{szymanska2015computational} and Martin \textit{et al.}\cite{martin2013computational}, and in Supplementary Fig.\ S2, we compare simulations of AV dynamics based on Eq.\ \eqref{eq:ode} with experimental measurements of AV dynamics reported by Martin \textit{et al.}\cite{martin2013computational}. Parameter settings are further explained in Supplementary Methods. In Supplementary Methods, we also elaborate on how earlier models\cite{szymanska2015computational,martin2013computational} guided our formulation of Eq.\ \eqref{eq:ode} and how these models differ from Eq.\ \eqref{eq:ode}.

Model-predicted physiological regulation of autophagy, by energy and nutrients, is summarized in Fig.~\ref{fig:bifur}. Figure \ref{fig:bifur}{\it A} shows how qualitative long-time behavior depends on the supplies of energy and nutrients, when these supplies are maintained at constant levels and in the absence of external control inputs $(u_i=0, i = 1,\ldots,6)$. Figures \ref{fig:bifur}{\it B--E} show time courses of autophagy induction or repression triggered by different energy/nutrient changes. All together, these plots show that model predictions of responses to physiological perturbations (i.e., changes in $C_{\rm En}$ and $C_{\rm Nu}$) are consistent with expectations based on the studies of Martin \textit{et al.}\cite{martin2013computational} and Szyma{\'{n}}ska \textit{et al.}\cite{szymanska2015computational}.

Dose-response curves predicted by the model for single-drug, constant-concentration perturbations are shown in Fig.~\ref{fig:constinput}. As can be seen, with increasing dosage, drugs 1 and 5 tend to increase the number of AVs per cell, whereas the other drugs tend to decrease the number of AVs per cell. These results are consistent with negative regulation of autophagy by MTORC1 and positive regulation of autophagy by ULK1, AMPK, and VPS34. As is the case for some physiological conditions (Fig.~\ref{fig:bifur}), AV count oscillates at some of the drug doses, depending on the supplies of energy and nutrients. All together, the plots shown in Fig.~\ref{fig:constinput} indicate that responses to single-drug, constant-concentration perturbations are consistent with accepted regulatory influences of MTORC1, ULK1, AMPK and VPS34 on autophagy.

As can be seen in Fig.~\ref{fig:constinput}, the ability of each drug $i$ to influence $x_5$ depends on the supplies of energy and nutrients, meaning the values of $C_{\rm En}$ and $C_{\rm Nu}$ (cf. the left and right panels in each row). In this figure, two energy/nutrient conditions are considered ($C_{\rm En}=C_{\rm Nu}=0.1$ and $0.6$); additional conditions are considered in Supplementary Figs. S3 and S4. Taken together, these results define the condition-dependent ranges over which $x_5$ can be feasibly controlled by each drug $i$.

\subsection*{Therapy design as an optimal control problem}

To design optimal therapies, we must first introduce design goals. Below, we introduce a series of goals/constraints that we will require optimal therapies to satisfy. However, let us first introduce notation useful for referring to therapies. We will refer to the set of six available drugs, or more precisely, drug types, as $\mathcal{D} =\{1,\ldots,6\}$, and we will refer to a therapy involving $k$ drugs chosen from $\mathcal{D}$ as $\mathcal{T}_k$, where
\begin{equation}\label{eq:therapy}
\begin{aligned}
\mathcal{T}_k \subseteq \mathcal{D} \quad \text{s.t.} \quad |\mathcal{T}_k | = k.
\end{aligned}
\end{equation}
Thus, for example, we will use $\mathcal{T}_1$ to refer to a monotherapy, and $\mathcal{T}_2$ to refer to a dual therapy. There are six possible monotherapies and, in general, $C^6_k$ distinct therapies that combine $k$ of the six drugs. Here, we will focus on monotherapies and dual therapies, leaving the evaluation of higher-order combination therapies for future work. As a simplification, we will assume that drugs used together in a combination do not interact. Thus, for example, for dual therapy with drugs 2 and 6 (Fig.~\ref{fig:model}), we consider these drugs to bind/inhibit VPS34 independently (i.e., non-competitively).

Our first, and most important, therapy design goal can be described (somewhat informally) as follows. Starting from a stationary (or recurrent) state at time $t=0$, we wish to use drug injections (i.e., drug inputs) according to a schedule defined by ${\bf u}(t)=(u_1(t),\ldots,u_6(t))$ to eventually maintain, after a transient of duration $t_0$, the number of AVs in an average cell, $x_5$, near (to within a tolerance $\epsilon$) a specified target level, $x_5^f$, for a period of at least $t_f-t_0$ ($t_f>t_0>0$), thereby achieving sustained control of the level of autophagic degradative flux in a cell, which is given by $k_4x_5$ according to Eq.~\eqref{eq:ode}. In our analyses, we will consider $t_0 = 120$ min and $t_f = 240$ min because these times are longer than typical transients (Figs.~\ref{fig:bifur}\emph{B}--\emph{E}).

A second therapy design goal of interest is minimization of the total amount of drug used, which is motivated by a desire to avoid drug toxicity arising from dose-dependent offtarget effects. In the optimal control literature, a problem entailing this type of constraint is called a \emph{minimum fuel} problem \cite{kirk2012optimal,lewis2012optimal}. The constraint can be expressed mathematically as follows:
\begin{equation}\label{eq:obj}
\begin{aligned}
\min_{\substack{u_i(t), \\ i \in \mathcal{T}_k}} J\left\{u_i\right\}:=   \sum_{ i \in \mathcal{T}_k} \int_{0}^{t_f} u_i(t)  \dd t
\end{aligned}
\end{equation}
where $u_i(t) \geq 0$ for $i = 1,\ldots,6$. As a simplification, we are considering an objective functional $J\left\{u_i\right\}$ that treats the different drugs equally, i.e., the sum in Eq.\ \eqref{eq:obj} is unweighted. With this approach, we are assuming that the different drugs of interest have equivalent toxicities. If drugs are known to have different toxicities, this assumption can be lifted simply by introducing weights to capture the toxicity differences, with greater weight assigned for greater toxicity. Indeed, arbitrary modifications of the form of the objective functional $J\left\{u_i\right\}$ would be feasible if such modifications are needed to capture problem-specific constraints on drug dosing.

A third design goal is to disallow the instantaneous concentration of any drug $i$, $w_i(t)$, from ever rising above a threshold $w_i^{\max}$. The rationale for this constraint is again related to a desire to eliminate or minimize dose-dependent drug toxicity. In other words, we are assuming that a drug $i$ is tolerable so long as its concentration $w_i$ is below a toxicity threshold $w_i^{\max}$. In our analyses, we set the toxicity threshold of a drug as a factor ($>1$) times its EC$_{50}$ dosage, which we define as the concentration of the drug at which its effect on $x_5$, negative or positive, is half maximal (see Eqs.\ \eqref{eq:Hill_h} and \eqref{eq:Hill_H}).

We are now prepared to formulate the problem of (combination) therapy design as a constrained, optimal control problem. The problem, for a given $\mathcal{T}_k$ (Eq.\ \eqref{eq:therapy}), is to find a drug schedule ${\bf u}(t)$ that minimizes the objective functional defined in Eq.\ \eqref{eq:obj} and that also satisfies the following constraints:
\begin{subequations}\label{eq:const}
	\begin{align}
	\dot{\textbf{X}}(t) ={}& \textbf{f}(\textbf{X}(t), \textbf{u}(t)), \quad 0\le t \le t_f, \\
	b_i  ={}& \begin{cases}
	1, & \text{if $i \in \mathcal{T}_k$},\\
	0, & \text{otherwise},
	\end{cases} \\
	x_5^f-\epsilon \le{}& x_5(t) \le x_5^f+\epsilon, \quad t_0\le t \le t_f, \label{eq:tube}\\
	0 \le{}& w_i(t) \le w_i^{\max}, \quad i = 1,\ldots,6, \label{eq:upperBound} \\
	0 \le{}& u_i(t),\quad i = 1,\ldots,6,	\\
	\textbf{X}(0) ={}& [\textbf{x}(0),\textbf{w}(0)]\equiv[\textbf{x}_0 , \textbf{0}].
	\end{align}
\end{subequations}
Here, $\textbf{X}(t)$ is defined as $[\textbf{x}(t), \textbf{w}(t)]$, where $\textbf{x}(t)=(x_1(t),\ldots,x_5(t))$ and $\textbf{w}(t)=(w_1(t),\ldots,w_6(t))$, and $\textbf{f}(\textbf{X}(t),\textbf{u}(t))$ is the vector field of Eq.\ \eqref{eq:ode}. The initial condition $\textbf{X}_0 = \textbf{X}(0)$ is taken to be a stationary (or recurrent) state of Eq.\ \eqref{eq:ode} where supplies of energy and nutrients are constant (i.e., $C_{\rm En}$ and $C_{\rm Nu}$ are fixed) and drugs are absent (i.e., $\textbf{u}(t)=0$). With this formulation, it should be noted that we are attempting to drive the system variable $x_5$ to a specified final value $x_5^f$ (to within a tolerance $\epsilon$), but we are making no attempt to control the other system variables $x_1$, $x_2$, $x_3$, and $x_4$. This approach is called target control  \cite{klickstein2017energy,shirin2017optimal}. In all of our analyses, we set $\epsilon=1$.

A useful measure of the amount of `fuel' used to achieve drug control of autophagy is the total dosage of drug $i$ used up to time $t$ during a therapy $\mathcal{T}_k$, which we denote as $r^\ast_{i,k}(t)$. This quantity is calculated using
\begin{equation} \label{eq:rcum}
r^\ast_{i,k}\left(t\right) = \int_{0}^{t}  u^\ast_i(\tau) d \tau,
\end{equation}
where $u^\ast_i(t)$ for $i \in \mathcal{T}_k$ is the solution of the nonlinear optimal control problem defined by Eqs.\ \eqref{eq:obj} and \eqref{eq:const}.  

\subsection*{Optimal monotherapies}

We will illustrate generic features of solutions to the nonlinear optimal control problem defined by Eqs.\ \eqref{eq:obj} and \eqref{eq:const} by focusing on a particular (severe) energy/nutrient stress condition (i.e., the condition where $C_\text{Nu}=C_\text{En}=0.1$). For this condition, the system represented by Eq.\ \eqref{eq:ode} has a near maximal, steady-state AV count of approximately 37 per cell (i.e., $x_5 \approx 37$). Let us focus for the moment on monotherapy with drug $4$ (an AMPK inhibitor) to downregulate the number of AVs to a target level of 10 per cell (i.e., $x_5^f = 10$) over the time period between $t_0=120$ min and $t_f=240$ min from an unperturbed steady state (i.e., dynamics with $u_i=0$) at $t=0$.

We solved the optimal control problem using the approach outlined in the Methods section and described in more detail in ``Pseudo-Spectral Optimal Control'' in Supplementary Methods. The solution, represented by the optimal cumulative dosage of drug $4$ (i.e., $r^\ast_{4,1}\left(t\right)$) (Eq.\ \eqref{eq:rcum}), is presented in Fig.~\ref{fig:best_mono}\emph{A}. The optimal solution exhibits several generic features of the system's dynamics, regardless of its parameterization. First, the computation suggests an optimal earliest time to apply the drug. In this particular example, this time is $t\lesssim 60$ min.  The difference between the target time $t_0$ and the earliest time to apply the drug quantitatively measures the speed of action of the drug. Secondly, the function $r^\ast_{4,1}\left(t\right)$ exhibits a staircase behavior, indicating that the optimal strategy of drug administration for this particular problem is to intermittently inject a specific dosage of drug into the system at specific times. Mathematically, this is due to the fact that the objective functional (Eq.\ \eqref{eq:obj}) is a linear combination of the $L^1$ norm of the injection/input rate $u_i$'s---see Sections 5.5 and 5.6 in Kirk\cite{kirk2012optimal}. 

Figure \ref{fig:best_mono}\emph{B} depicts how the drug concentration $w_4(t)$ evolves subject to the optimal protocol $u^\ast_4(t)$.
We observe surges of $w_4(t)$ in response to the drug being applied to the system in large quantities over small intervals, and slow decays in between applications of the drug (caused by the natural decay of the drug concentration in the absence of external drug inputs dictated by $\delta_i$.) As a consequence, the optimal solution is to inject a relatively large dose of a drug periodically, and to continuously supply small amounts of that drug to replenish drug cleared from the system to stably maintain autophagic flux (i.e., constant AV count and constant degradative flux, which we take to be proportional to the AV count). 

Figure \ref{fig:best_mono}\emph{C} illustrates the time evolution of $x_5$ (AV count) subject to the optimal drug administration protocol.
As can be seen, for $t \geq 120$ min, $x_5$ is maintained within the desired interval $x_{5_f} \pm \epsilon=10 \pm 1$. The time evolution of the non-target variables $x_1$, $x_2$, $x_3$ and $x_4$ (i.e., the activities of the regulatory kinases) are presented in Fig.~\ref{fig:best_mono}\emph{D}. Together, Figs.~4\emph{C} and \emph{D} provide a full representation of the time evolution of the system represented by Eq.\ \eqref{eq:ode} (the target and non-target variables) under the influence of the optimal drug administration schedule. Because our procedure to find the optimal solution to the nonlinear optimal control problem is numerical, we have verified that the optimal control solution satisfies the necessary conditions that it must satisfy for optimality. See ``Pseudo-Spectral Optimal Control'' in Supplementary Methods for details. 

Given that cancer cells may be killed by using drugs to either elevate or suppress autophagy \cite{MulcahyLevy2017}, we will now consider optimal control solutions that either upregulate or downregulate autophagic flux by using a single drug. We will identify the drugs which can perturb and maintain the system near the target AV count. Perhaps more importantly, our analysis will deliver optimal protocols which include the precise times to inject the drugs, whose dosages are also tightly controlled to minimize the total quantities of drugs that are supplied. 

\begin{figure}[t!] 
	\centering
	\includegraphics[width = 0.48\textwidth]{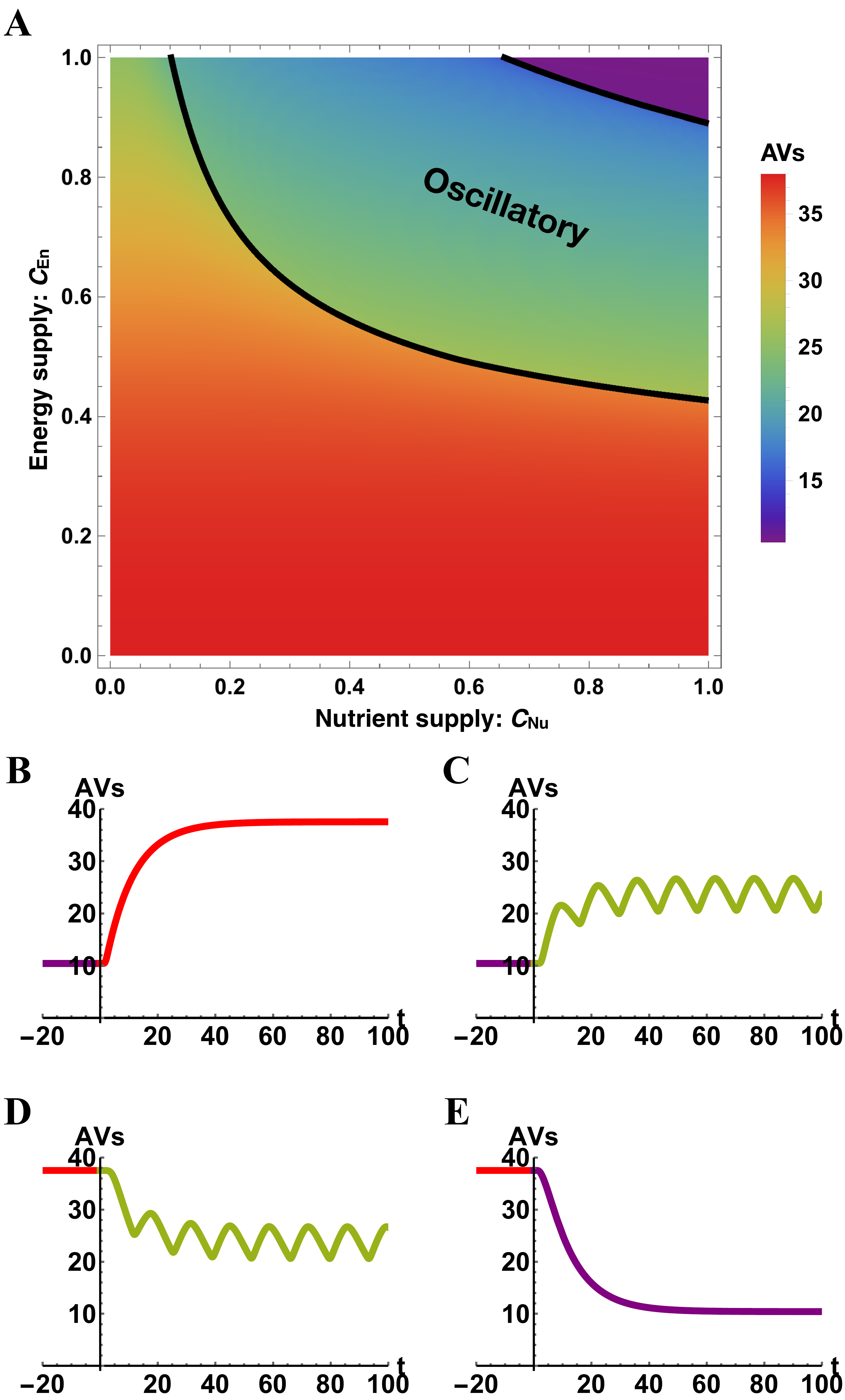}
	\caption{Predicted dependence of AV count on energy and nutrient supplies according to the model for autophagy regulation (Eq.\ \eqref{eq:ode}). (\emph{A}) Long-time behavior. In this panel, the stationary or time-averaged value of $x_5(t)$
	for constant supplies of energy and nutrients as $t \rightarrow \infty$ is indicated by color over the full ranges of the two physiological inputs of the model: energy supply ($C_{\rm En}$) and nutrient supply ($C_{\rm Nu}$). It should be noted that we take the most extreme energy/nutrient starvation conditions to correspond to $C_{\rm En}=C_{\rm Nu}=0$, and we take the most extreme energy/nutrient replete conditions to correspond to $C_{\rm En}=C_{\rm Nu}=1$. The solid black curves delimit the regions where long-time behavior of $x_5$ is oscillatory or not. If behavior is oscillatory, the time-averaged value of $x_5$ is reported; otherwise, the stationary value is reported. A bifurcation analysis indicates that long-time behavior is characterized by a stable fixed point, the coexistence of a stable fixed point \emph{and} a stable limit cycle, or a stable limit cycle. The region labeled `oscillatory' indicates the conditions for which a stable limit cycle exists; however, this diagram is not intended to provide a full characterization of the possible qualitative behaviors and bifurcations of Eq.\ \eqref{eq:ode}. As indicated by the color bar, the (average) AV count varies over a range of roughly 2 to 37 vesicles per cell. (\emph{B}--\emph{E}) Transient behavior. Each of these plots shows $x_5$ as a function of time $t$ after a coordinated change in energy and nutrient supplies. The plot in panel \emph{B} shows the predicted response to a steep, step increase in stress level, i.e., a change in conditions from $C_{\rm En}=C_{\rm Nu}=1$ to $0.2$. The plot in panel \emph{C} shows the predicted response to a moderate, step increase in stress level, i.e., a change in conditions from $C_{\rm En}=C_{\rm Nu}=1$ to $0.6$. The plot in panel \emph{D}  shows the predicted response to a moderate, step decrease in stress level, i.e., a change in conditions from $C_{\rm En}=C_{\rm Nu}=0.2$ to $0.6$ The plot in panel \emph{E} shows the predicted response to a steep, step decrease in stress level, i.e., a change in conditions from $C_{\rm En}=C_{\rm Nu}=0.2$ to $1$.}
	\label{fig:bifur}
\end{figure}

\begin{figure}[t!]
	\centering
	\includegraphics[width=0.48\textwidth]{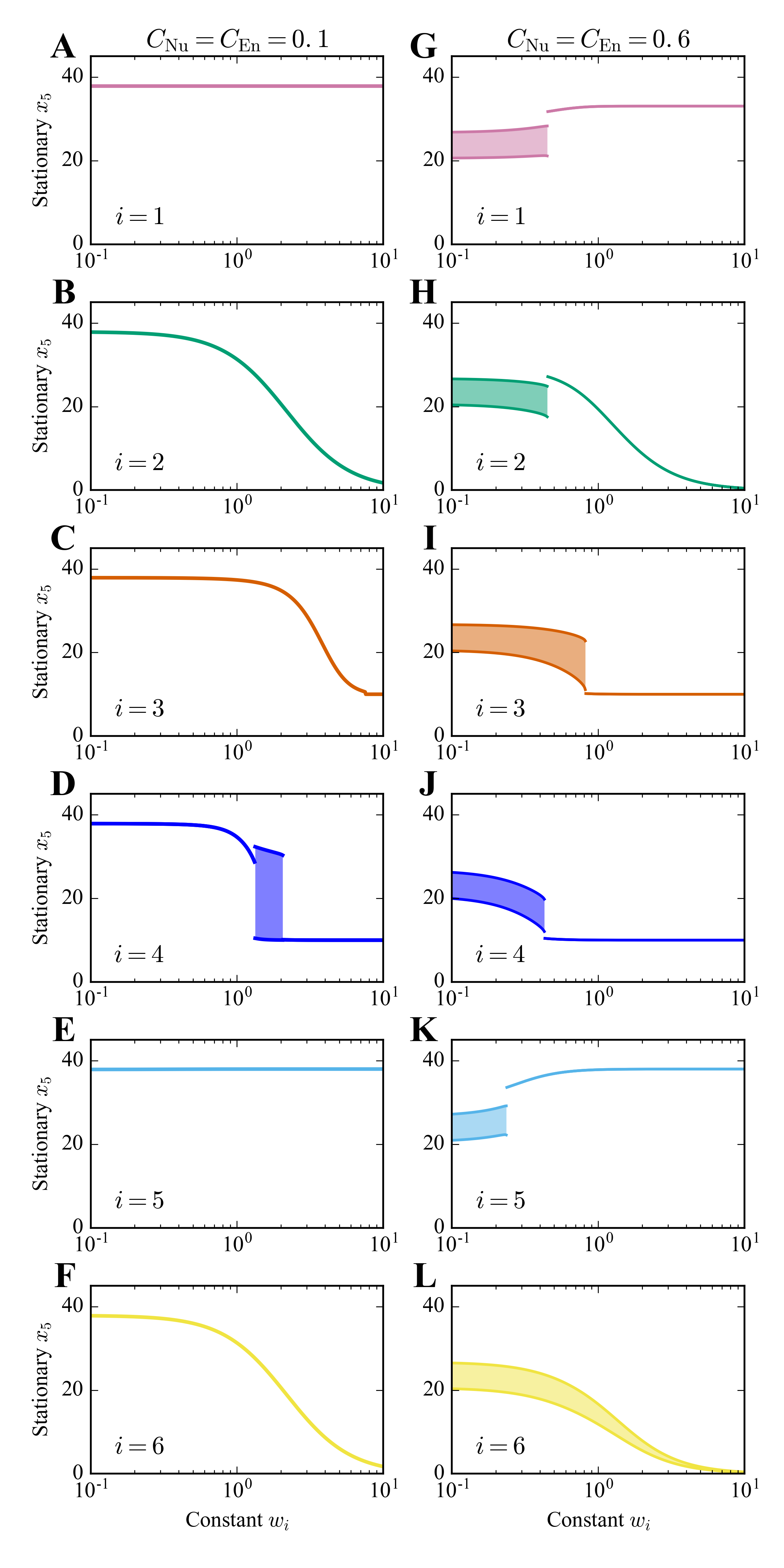}
	\caption{Predicted dependence of AV count ($x_5$) on drug dose according to Eq.\ \eqref{eq:ode}. In each panel, we show the long-time effects of monotherapy with drug $i \in \{1,\ldots,6\}$; the drug considered in each panel is maintained at the constant (dimensionless) concentration indicated on the horizontal axis. Drugs 1--6 are considered from top to bottom. Responses to drugs depend on the supplies of energy and nutrients. The left panels (\emph{A}--\emph{F}) correspond to conditions for which $C_\text{Nu} = C_\text{En}=0.1$ (severe energy/nutrient stress), and the right panels (\emph{G}--\emph{L}) correspond to conditions for which $C_\text{Nu} = C_\text{En}=0.6$ (moderate energy/nutrient stress). The long-time behavior of $x_5$ under the influence of monotherapy can be stationary (with a stable fixed point) or oscillatory (with a stable limit cycle). The shaded regions indicate where there is oscillatory behavior. At a given drug dose, the top and bottom bounds of a shaded region delimit the envelope of oscillations (i.e., the maximum and minimum values of $x_5$).}
	\label{fig:constinput}
\end{figure}

\begin{figure}[t!]
	\centering
	\includegraphics[width=17.8cm]{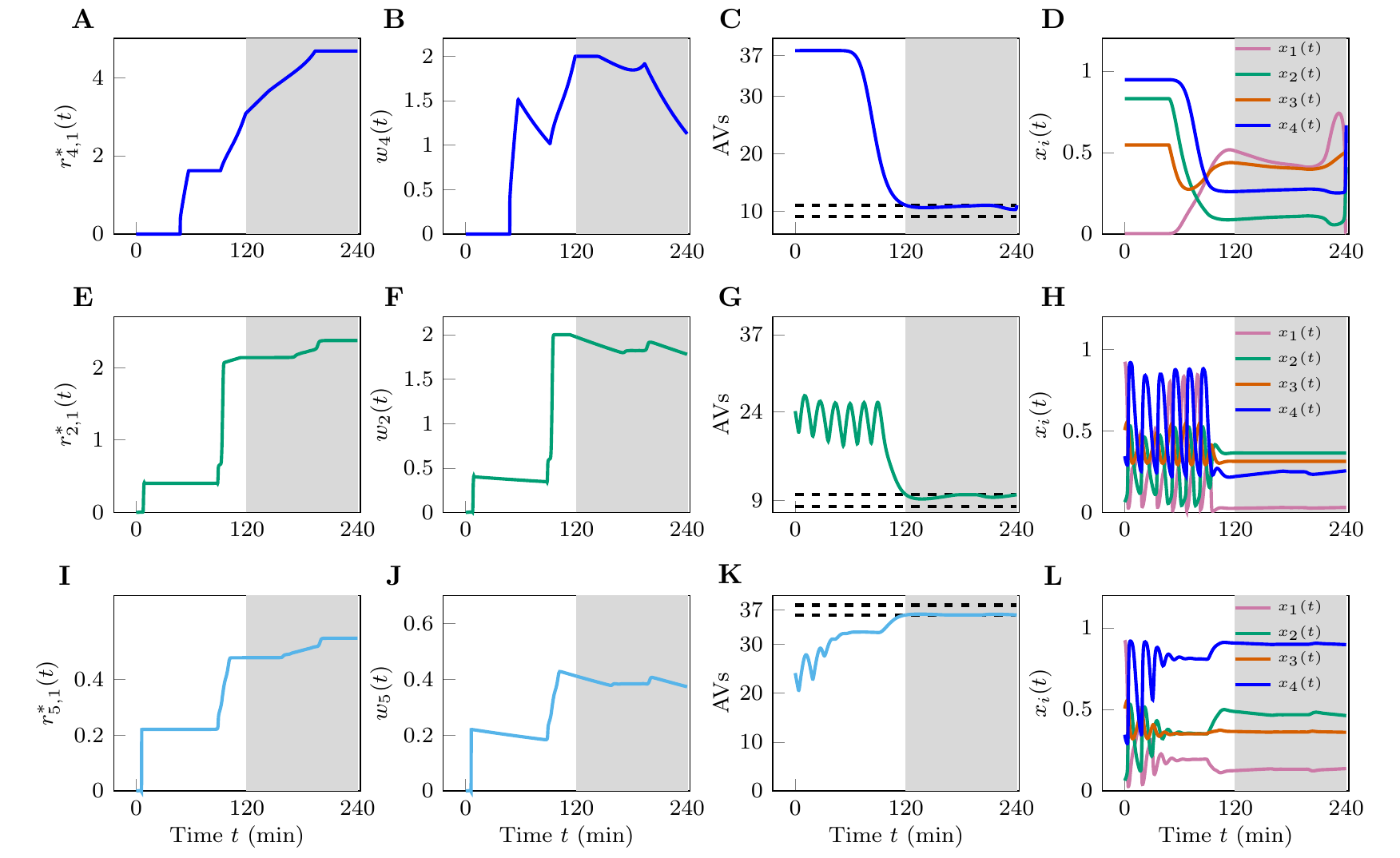}
	\caption{Best performing monotherapies.  
		(\emph{A}--\emph{D}) Panels \emph{A}--\emph{D} are from a numerical experiment for which we set $C_{\text{Nu}} = C_{\text{En}}= 0.1$ and attempt to use drug 4 to downregulate the AV count. (\emph{E}--\emph{H}) Panels \emph{E}--\emph{H} from a numerical experiment for which we set $C_{\text{Nu}} = C_{\text{En}}= 0.6$ and attempt to use drug 2 to downregulate the AV count. (\emph{I}--\emph{L}) Panels \emph{I}--\emph{L} are from a numerical experiment for which we set $C_{\text{Nu}} = C_{\text{En}}= 0.6$ and attempt to use drug 5 to upregulate the AV count. The plots in the first column are cumulative drug dosages for the monotherapies considered. The plots in the second column are the drug concentrations. The plots in the third column show $x_5(t)$ and the plots in the fourth, or rightmost, column show $x_1(t)$, $x_2(t)$, $x_3(t)$, and $x_4(t)$ that we are making no attempt to control. In all simulations, the upper bound on the allowable concentration of drug $i$, $w_i^{\max}$, was set at $2$. For panels \emph{A}--\emph{H}, the target AV count was 10 (i.e., $x_5^f=10$). For panels \emph{I}--\emph{L}, the target AV count was 37 (i.e., $x_5^f=37$). The white region corresponds to the time interval $[t_0,t_f]$ when we either upregulate or downregulate the AV count The shaded region corresponds to the time interval $[t_0, t_f]$ when the AV count is maintained within the interval $x_5^f \pm \epsilon$.}
	\label{fig:best_mono}
\end{figure}

Let us consider the case of intermediate energy/nutrient stress before treatment (i.e., the condition corresponding to $C_\text{Nu}=C_\text{En}=0.6$; see Fig.~\ref{fig:bifur}), for which the system exhibits oscillations in the range $[20,27]$ without treatments. For this scenario, our goal is to either downregulate the number of AVs to $x_5^f \approx 9$ (shown in Figs.~\ref{fig:best_mono}\emph{E}--\emph{H}) or to upregulate the AVs to $x_5^f\approx 37$ (shown in Figs.~\ref{fig:best_mono}\emph{I}--\emph{L}). We have performed extensive numerical solutions of the monotherapy optimal control problem with various settings of the parameters $w_i^{\max}$, $t_0$, $t_f$ and $x_5^f$.
We set the control window in the interval between $t_0=120$ min and $t_f=240$ min and imposed a constraint on each drug concentration $w_i$, requiring it not to exceed $w^{\max} = 4 \times {\rm EC}_{50}$.  

We found drug 2 to be best suited for downregulation for two reasons. First, drug 2 is able to drive $x_5$ nearly to zero (in contrast with the case for drug 3 or 4). See Figs.~\ref{fig:bifur}\emph{B} and \ref{fig:bifur}\emph{H} and compare with Figs. \ref{fig:bifur}\emph{C}, \ref{fig:bifur}\emph{D}, \ref{fig:bifur}\emph{I}, and \ref{fig:bifur}\emph{J}. Second, drug 2 (in contrast with drug 6) is able to overcome the autonomous oscillatory behavior in $x_5$. In the analysis summarized in Supplementary Fig. S7, we found that drug 6 cannot eliminate oscillatory behavior; thus, it is incapable of maintaining a low, steady AV level. Drug 6 becomes viable if we remove the lower bound from the constraint of Eq.\ \eqref{eq:tube}. Without the lower bound, oscillations in $x_5$ are permitted. We choose to keep the constraint of Eq.\ \eqref{eq:tube} as written to avoid oscillatory solutions because, depending on period and amplitude, oscillations in $x_5$ may allow for autophagy-addicted cells to survive periods of relatively low autophagy by thriving during periods of relatively high autophagy. In the other direction (i.e., drug-induced upregulation of autophagy), it is only possible to use drug 5 to upregulate autophagy to the target value $x_5^f=37$ (Fig.~\ref{fig:constinput}). Figs.~\ref{fig:best_mono}\emph{E}--\emph{H} and \ref{fig:best_mono}\emph{I}--\emph{L} illustrate the optimal solutions using drugs 2 and 5 to downregulate and upregulate autophagy, respectively.

Although the selection of a single drug to achieve a given qualitative change in $x_5$ is intuitive, especially given the results of Fig.~\ref{fig:constinput}, optimization of drug scheduling (Fig.~\ref{fig:best_mono}) delivers better solutions in the sense that the total dosage applied to achieve the same effect (compared to constant input) is lower (minimized). Furthermore, the generic staircase-like solutions for $r^\ast_{i,k}$ illustrated in Fig.~\ref{fig:best_mono} persist for all the parameter sets we have tested (see below), indicating that variable, tightly controlled dosages should be injected into the system at controlled times. Given a particular type of drug, the central result of our optimal control analysis is to provide injection/input times and the amounts of drugs to be injected/added.

\subsection*{Optimal combination therapies}

Let us now consider dual therapies ($k=2$). The motivation is to identify therapies---protocols involving lower quantities of drugs and faster responses---that are even more efficient than optimal monotherapies. We have evaluated all possible dual therapies ($C^6_2 = 15$) for each of two energy/nutrient stress conditions: $C_\text{En}=C_\text{Nu}=0.1$ (corresponding to severe stress) and $C_\text{En}=C_\text{Nu}=0.6$ (corresponding to moderate stress). With an identical control objective and identical constraints $w_i^{\max}=2.0$  $t_0=120$, $t_f=240$, $x_5^f=10$, and $\epsilon=1$, we found four pairs of drugs that are each more efficient than the optimal monotherapy with either of the two drugs included in the combination. These dual therapies are illustrated in Fig.~\ref{fig:best_dual}. Additional results from our analyses of dual therapies are presented in the Supplementary Note and Supplementary Figs. S3--S10. 

We found that when baseline autophagy is high ($C_\text{En}=C_\text{Nu}=0.1$), the only combination of drugs that can drive AV count down to the target $x_5^f$ is the combination of drugs 2 and 6. The dynamical response of the system is shown in Figs.~\ref{fig:best_dual}\emph{A}--\emph{D}. For this particular combination, either drug alone cannot lower $x_5$ to $10$ without violating one or both of the constraints $w_i< w_i^{\max}$ ($i=2$, $6$). However, with use of drugs 2 and 6 in combination, it is possible to achieve the target AV count because the effects of the drugs are multiplicative (Eq.\ \eqref{eq:drug2and6}) and drug $2$ directly affects both MTORC1 (Eq.\ \eqref{eq:drug1and2}) and VPS34 (Eq.\ \eqref{eq:drug2and6}). 

Our analysis predicts non-trivial synergistic activities between drugs when the baseline level of autophagy is intermediate (on average) and exhibits oscillatory behavior ($C_\text{En}=C_\text{Nu}=0.6$). The results are summarized in Figs.~\ref{fig:best_dual}\emph{E}--\emph{P}. In this scenario, multiple drug combinations (drugs 1 and 6, 2 and 6, and 3 and 6) are able to downregulate and stabilize $x_5$, whereas drug 6 alone cannot do so. Using drug 6 alone results in oscillations in $x_5$, causing a violation of the constraint of Eq.\ \eqref{eq:tube}. More interestingly, the optimal application of the drugs reveals a clear sequential protocol: first apply a drug other than drug 6 (1, 2, or 3) to suppress oscillations (see Figs.~\ref{fig:best_dual}\emph{H}, \emph{L} and \emph{P}), then  apply drug 6 to drive AV count down to the desired level. The combination of drugs 1 and 6 is peculiar in that in this case application of drug 1 drives the system out of the oscillatory regime (Fig.~\ref{fig:best_dual}\emph{O}) but also upregulates autophagy; subsequent application of drug 6 is effective in downregulating autophagy.

It is important to emphasize that the two drugs acting together in any given combination therapy are, for simplicity, modeled as non-interacting, which may or may not be reasonable, depending on the mechanisms of actions of specific drugs of interest. The drug synergies detected in our analyses arise from the nonlinear dynamics of the regulatory network controlling autophagy. Without the formal framework presented here for therapy design, it would arguably be difficult to identify these synergies.



\section*{Discussion}

Here, we have taken up the problem of designing targeted therapies to control a cellular phenotype of cancer cells, namely, their commitment to recycling of cytoplasmic contents through the process of autophagy, as measured by cellular autophagic vesicle (AV) count. Autophagy generates building blocks needed for \textit{de novo} protein synthesis in support of growth (and proliferation). Modulation of autophagy, up or down, in autophagy-addicted cancer cells has the potential to selectively kill these cells \cite{MulcahyLevy2017}.

Our approach was to first construct a mathematical model for autophagy regulation that captures the effects of key physiological stimuli---changes in the supplies of energy and nutrients---and the idealized effects of six available drug types (Eq.\ \eqref{eq:ode}, Figs. \ref{fig:model}--\ref{fig:constinput}) and to then pose the question of therapy design as a constrained, optimal control problem (Eqs.\ \eqref{eq:therapy}--\eqref{eq:const}). Numerical solution of this problem, through optimization of a control input accounted for in the model (i.e., an adjustable time-dependent drug injection/input rate), yielded monotherapy drug schedules that require a minimum amount of drug, maintain drug concentration below a specified threshold at all times, and that bring about desired effects in the most efficient manner possible (Fig.~\ref{fig:best_mono}), in a well-defined sense. Furthermore, through the essentially same approach, but with consideration of adjustable time-dependent drug injection/input rates for two different drugs, we were able to predict synergistic drug pairs (Fig.~\ref{fig:best_dual}).

Optimal monotherapies were found to entail intermittent pulses of drug injection/input at irregular, non-obvious intervals and doses (Fig.~\ref{fig:best_mono}). These features of optimal drug schedules---the pulsatile nature of drug administration and the irregularity of drug administration in terms of both timing and dosage---appear to be generic and each is discussed in further detail below.

The pulsatile nature of optimal monotherapy arises from the optimal control problem that we posed (Eqs.\ \eqref{eq:therapy}--\eqref{eq:const}), which can be viewed as a minimum-fuel problem, in that our control problem calls for usage of a minimal total amount of drug. The rationale for this control objective is that drugs typically have dose-dependent offtarget effects, which may contribute to drug toxicity. Thus, by seeking drug schedules that achieve desired endpoints while using only a minimal total amount of drug, we seek to mitigate the possible negative consequences of offtarget drug effects. Mathematically, our minimum-fuel objective function, Eq.\ \eqref{eq:obj}, leads to pulsatile drug administration because the Hamiltonian of the optimal control problem is linear in the control inputs $u_i(t)$, $i \in \mathcal{T}_k$ (see ``Pseudo-Spectral Optimal Control'' in Supplementary Methods for a detailed derivation). Optimal control problems which have Hamiltonians that are linear in the control input are well-known to have singular arcs, that is, discontinuities jumping between upper and lower bounds of the control input (see Chapter 5 in Kirk\cite{kirk2012optimal} for the derivation of singular arc behavior and the brief overview of this issue in ``Pseudo-Spectral Optimal Control'' in Supplementary Methods). Because we do not impose an upper bound on $u_i(t)$, the discontinuities we expect to see are Dirac delta type functions, a pulse of infinite magnitude but infinitesimal width. With the use of numerical methods to find solutions of the optimal control problem, we cannot capture the Dirac delta behavior exactly.
Instead, we see finite pulses of finite width, which, while likely suboptimal, are more physically realistic.

Although pulses of drug input are consistent with convenient drug delivery modalities, such as oral administration of a drug in pill form or intravenous injection, optimal schedules do not entail uniform drug doses, nor uniform periods of drug administration. This irregular nature of optimal drug administration depends on the structure of the nonlinear cellular network that controls the synthesis of AVs. In particular, in our model, each drug specifically targets individual nodes of the cellular network, and therefore, different drugs play dynamically distinct roles and cannot be treated as equivalent control inputs. Thus, it may be critically important to better understand the interplay between targeted therapies and archetypical cellular regulatory network dynamics if we are to design the best possible therapies for populations of patients. Because network dynamics can be expected to vary between patients, patient-specific variability in network dynamics, which we have not considered in our analyses here, is a factor that likely affects the efficacy of individualized targeted therapy and that therefore should receive attention in future studies. The study of Fey \textit{et al.}\cite{fey2015signaling} points to the feasibility of considering patient-specific parameters in mathematical models. In this study, gene expression data available for individual patients were used to set the abundances of gene products in patient-specific models for a cell signaling system. Because mutations can be detected in the tumors of individual patients, effects of oncogenic mutations could also potentially be accounted for in patient-specific models. The study of Rukhlenko \textit{et al.}\cite{Rukhlenko2018} provides an example of a study where the effects of an oncogenic mutation were considered in a mathematical model. In the study of Fr{\"o}hlich \textit{et al.}\cite{frohlich2018efficient}, gene expression and mutational profiles were both considered in cell line-specific models.

The therapy design approach presented here is flexible and allows for the evaluation of drug combinations. In our analyses, we focused on dual therapies. Somewhat surprisingly, we found several drug pairs that together are more effective than either drug alone (according to our model). These pairs are drug 2 and drug 6 when $C_{\text{Nu}}=C_{\text{En}}=0.1$ (severe energy/nutrient stress) and the combination of drug 6 with drug 1, 2, or 3 when $C_{\text{Nu}}=C_{\text{En}}=0.6$ (moderate energy/nutrient stress). In the latter cases, drug 6 alone is incapable of downregulating autophagy to the desired level, but it sensitizes the network to drugs 1--3 when one of these drugs is used in conjunction with drug 6. According to the model (and its parameterization), the most potent synergistic drug pair is the combination of drugs 2 and 6. With this combination, the total amount of drug 2 used was reduced by more than 5-fold (see the Supplementary Note and Supplementary Fig. S5) in comparison to the case where drug 2 is used optimally in isolation. More striking perhaps is that drug 6 when used alone is incapable of achieving the performance objective. Interestingly, our results provide mechanistic insight into the optimal sequence of drug delivery: therapy is optimal when drug 2 is injected about 80 minutes earlier than drug 6. That is, the best outcome was achieved when first inhibiting MTORC1, thus halting the intrinsic oscillations of the network dynamics, and then only inhibiting VPS34 to reduce synthesis of AVs. It should be noted that in our evaluation of this drug pair, we have assumed that there is no interaction between drugs 2 and 6, an idealization that may not be appropriate for specific examples of drugs of these types. 

The same optimal control approach that we have demonstrated for 2-drug combinations can be applied for combinations involving more than two drugs. Indeed, our approach was presented for the general case of $k$ drugs used in combination. Our expectation is that effective combinations involving more than two drugs may be more likely to exist than effective combinations involving only two drugs, because controllability would presumably increase with the availability of more drugs. However, finding an effective combination may be more computationally expensive because of the larger number of possible combinations, and 2-drug combinations may be preferable to higher-order combinations because of drug side effects.

As reported by Palmer and Sorger \cite{Palmer2017}, many clinically used drug combinations are effective for reasons other than drug synergy, which is rare. In essence, the majority of clinically available drug combinations are, for all intents and purposes, equivalent to monotherapy at the level of individual patients. The basis for their effectiveness at the population level is simply that tumors in different subpopulations of patients have distinct drug sensitivities. Thus, new methods for predicting promising, non-obvious synergistic drug combinations, such as the approach reported here, could be helpful in developing combination therapies that derive their effectiveness from drug synergy. Synergistic drug combinations would seemingly offer significant benefits over monotherapy, or what is effectively monotherapy, in terms of delaying or perhaps eliminating the emergence of drug resistance. We note that our analysis identified synergies between pairs of drugs that are predicted to manifest without fine tuning of the doses used or the timing of drug administration. We admit that these predictions could perhaps have been found through an \textit{ad hoc} model analysis. Nevertheless, we see value in leveraging an optimal control framework for model analysis, even if an optimal control strategy is not sought, because with this type of approach it is less likely that interesting behavior will be missed.

There is presently cautious optimism that effective drug combinations will be identified through high-throughput screening experiments \cite{Holbeck2017}, or through learning from data. However, the sheer number of possible drug combinations poses a barrier to experimental discovery of efficacious drug combinations and it is not clear that the data requirements of machine learning approaches can be met in the near term. Thus, it is important to consider alternatives, such as the approach presented here, which leverages available mechanistic understanding of how regulatory protein/lipid kinases influence the synthesis of AVs, which we have consolidated in the form of a mathematical model (Eq.\ \eqref{eq:ode}), designed to be useful for computational characterization of drug combinations. We note that our model was formulated specifically for this purpose, and it was not designed to make predictions outside this limited domain of application. Indeed, to facilitate our computational analyses, the model was handcrafted to be as simple as possible while still reproducing key behaviors of more mechanistically detailed models \cite{martin2013computational,szymanska2015computational}. This approach was helpful in making calculations feasible. Unfortunately, to our best knowledge, there are no proven approaches for systematically and automatically deriving a suitable surrogate model for therapy design from a more detailed, mechanistic model of a cellular regulatory network. Pursuit of such a capability seems like an important subject of future research.

Our intent at the start of this study was to investigate how control engineering concepts might be introduced into formal therapy design. Thus, we have only attempted to demonstrate that our methodology is capable of generating interesting (and testable) predictions of effective drug schedules and drug combinations. Development of novel therapies will, of course, require experimental validation of candidate combinations, which is beyond the intended scope of the present study. Thus, we caution that our predictions of optimal drug schedules and synergistic drug combinations are only intended to demonstrate methodology. The merit of this methodology is not in reaching final conclusions but in prioritizing experimental efforts and thereby accelerating experimental validation of targeted therapies. Because kinase inhibitors of each type considered in our analysis are available for experimental characterization and autophagy is a cellular phenotype that can be readily assayed, as in the study of Martin \textit{et al.}\cite{martin2013computational} or du Toit \textit{et al.}\cite{dutoit2018measuring}, a logical next step would be to probe for the predicted drug synergies in cell line experiments. It might be especially interesting to evaluate a combination of an ULK1-specific inhibitor, such as ULK-101 \cite{martin2018potent}, and a VPS34-specific inhibitor, such as VPS34-IN1 \cite{bago2014characterization}. We predict that this combination will be synergistic, and the combination targets the two kinases considered in our analysis that are most proximal to the cellular machinery for producing autophagosomes. On the computational side, to increase confidence in predictions, sensitivity analysis techniques tailored for optimal control problems could be applied to characterize the robustness of predictions \cite{castillo2008sensitivity,malanowski1998sensitivity}, and experimental design techniques could be applied to aid in generating data useful for reducing parameter uncertainty \cite{hagen2013convergence,dehghannasiri2015efficient}. Several studies strongly support the potential value of formal therapy design \cite{Chmielecki2011,Chakrabarti2017,stein2018mathematical}, and the main contribution here is a new approach to this subject. Two important distinguishing features of this approach are 1) the consideration of a mathematical model for a cellular regulatory network that controls a cellular phenotype and 2) application of sophisticated methods from automatic control theory.

\section*{Methods}

\subsection*{Simulations}
Simulations were performed by numerical integration of the model ODEs. The parameter settings used in calculations are provided in the Supplementary Tables S1 and  S2.

\subsection*{Pseudo-Spectral Optimal Control}
Optimal control as a field of research combines aspects of dynamical systems, mathematical optimization and the calculus of variations \cite{kirk2012optimal}.
Together Eqs.\ \eqref{eq:obj} and \eqref{eq:const} form a constrained optimal control problem, which can generally be written as,
\begin{equation}\label{eq:OCP}
\begin{aligned}
\min_{\textbf{u}(t)} && &J(\textbf{x}(t),\textbf{u}(t),t) = \int_{t_0}^{t_f} F\left( \textbf{x}(t), \textbf{u}(t), t\right) \dd t\\
\text{s.t.} && &\dot{\textbf{x}}(t) = \textbf{f} ( \textbf{x}(t), \textbf{u}(t), t)\\
&& &\textbf{e}^L \leq \textbf{e}(\textbf{x}(t_0), \textbf{x}(t_f), t_0, t_f) \leq \textbf{e}^U\\
&& &\textbf{h}^L \leq \textbf{h}(\textbf{x}(t), \textbf{u}(t), t) \leq \textbf{h}^U\\
&& &t \in [t_0,t_f]
\end{aligned}
\end{equation}
In general, there exists no analytic framework that is able to provide the optimal time traces of the controls and the states in Eq.\ \eqref{eq:OCP}, and so we must resort to numerical techniques.

Pseudo-spectral optimal control (PSOC) has become a popular tool in recent years \cite{rao2009survey, ross2012review} that has allowed scientists and engineers solve optimal control problems like that of Eq.\ \eqref{eq:OCP} reliably and efficiently in applications such as guiding autonomous vehicles and maneuvering the international space station \cite{ross2012review}.
The main concepts of PSOC are summarized here but are explained at length in ``Pseudo-Spectral Optimal Control'' in Supplementary Methods. See also Supplementary Fig. S11.
We define a set of $N$ discrete times $\{\tau_i\}$ $i = 0,1,\ldots,N$ where $\tau_0 = -1$ and $\tau_N = 1$ with a mapping between $t \in [t_0,t_f]$ and $\tau \in [-1,1]$. The choice of $\{\tau_i\}$ is key to the convergence of the full discretized problem and so typically they are chosen as the roots of an $N+1$th order orthogonal polynomial such as Legendre or Chebyshev.
In fact, the type of PSOC one uses is typically named after the type of polynomial used to generate the discretization points.
Let $\hat{\textbf{x}}(\tau) = \sum_{i=0}^N \hat{\textbf{x}}_i L_i(\tau)$ be an approximation of $\textbf{x}(\tau)$ where $L_i(\tau)$ is the $i$th Lagrange interpolating polynomial. The dynamical system is approximated by differentiating the approximation $\hat{\textbf{x}}(\tau) = \sum_{i=0}^N \hat{\textbf{x}}_i L_i(\tau)$ with respect to time.
\begin{equation}
\begin{aligned}
\frac{d \hat{\textbf{x}}}{d \tau} = \sum_{i=0}^N \textbf{x}_i \frac{d L_i}{d\tau}
\end{aligned}
\end{equation}
Let $D_{k,i} = \frac{d}{d\tau} L_i(\tau_k)$ so that we may rewrite the original dynamical system constraint in Eq.\ \eqref{eq:OCP} as the following set of algebraic constraints.
\begin{equation}
\sum_{i=0}^N D_{k,i} \textbf{x}_i - \frac{t_f-t_0}{2} \textbf{f}(\hat{\textbf{x}}_i,\hat{\textbf{u}}_i,\tau_i) = \boldsymbol{0},\ k = 1,\ldots,N
\end{equation}
 
With the original time-varying states and control inputs now discretized, the dynamical equations approximated with Lagrange interpolating polynomials, and the cost function approximated by a quadrature, the discretized optimal control problem can be expressed as the following nonlinear programming (NLP) problem.
\begin{equation}\label{eq:dOCP}
\begin{aligned}
\min_{\substack{\textbf{u}_i\\ i=0,\ldots,N}} && &\hat{J} = \frac{t_f-t_0}{2} \sum_{i=0}^N w_i f(\hat{\textbf{x}}_i,\hat{\textbf{u}}_i,\tau_i)\\
\text{s.t.} && &\sum_{i=0}^N D_{k,i} \hat{\textbf{x}}_i - \frac{t_f-t_0}{2} \textbf{f}(\hat{\textbf{x}}_k,\hat{\textbf{u}}_k,\tau_k) = \boldsymbol{0},\ k = 0,\ldots,N\\
&& &\textbf{e}^L \leq \textbf{e}(\hat{\textbf{x}}_0,\hat{\textbf{x}}_N,\tau_0,\tau_N) \leq \textbf{e}^U\\
&& &\textbf{h}^L \leq \textbf{h}(\hat{\textbf{x}}_k,\hat{\textbf{u}}_k, \tau_k) \leq \textbf{h}^U,\ k = 0,\ldots,N\\
&& &t_i = \frac{t_f-t_0}{2}\tau_i + \frac{t_f+t_0}{2}
\end{aligned}
\end{equation}
We used $\mathcal{PSOPT}$ \cite{becerra2010solving}, an open-source PSOC toolbox written in C++, to perform the PSOC discretization procedure.

The NLP problem of Eq.\ \eqref{eq:dOCP} can be solved with a number of different techniques, but here we use an interior point algorithm \cite{nocedal2006numerical} as implemented in the open-source C++ software Ipopt \cite{wachter2006implementation}.

\section*{Acknowledgements}

I.S.K., A.S. and F.S. acknowledge support from the National Science Foundation (CRISP-1541148), the Office of Naval Research (N00014-16-1-2637), and the Defense Threat Reduction Agency (HDTRA1-12-1-0020). W.S.H. and Y.T.L. acknowledge support from the National Cancer Institute of the National Institutes of Health (R01CA197398). S.F. acknowledges support from the Center for Nonlinear Studies and the Laboratory-Directed Research and Development program at Los Alamos National Laboratory, which is operated by Triad National Security, LLC for the National Nuclear Security Administration of the U.S. Department of Energy (contract no. 89233218CNA000001). W.S.H. performed part of this work at the Aspen Center for Physics, which is supported by the National Science Foundation (PHY-1607611). We thank Jeffrey P. MacKeigan for helpful discussions.

\section*{Author Contributions Statement}

W.S.H. and F.S. designed the research; A.S., I.S.K., S.F. and Y.T.L. performed the research; and all authors contributed to the analyses of results and the writing of the manuscript. Modeling work was the primary responsibility of the Los Alamos National Laboratory authors. Optimal control work was the primary responsibility of the University of New Mexico authors.

\section*{Additional Information}

The authors declare no competing interests.

\section*{Data availability}
Problem-specific software used in this study is provided as Supplementary Data.


\begin{figure}[tbhp!]
	\centering
	\includegraphics[width=17.8cm]{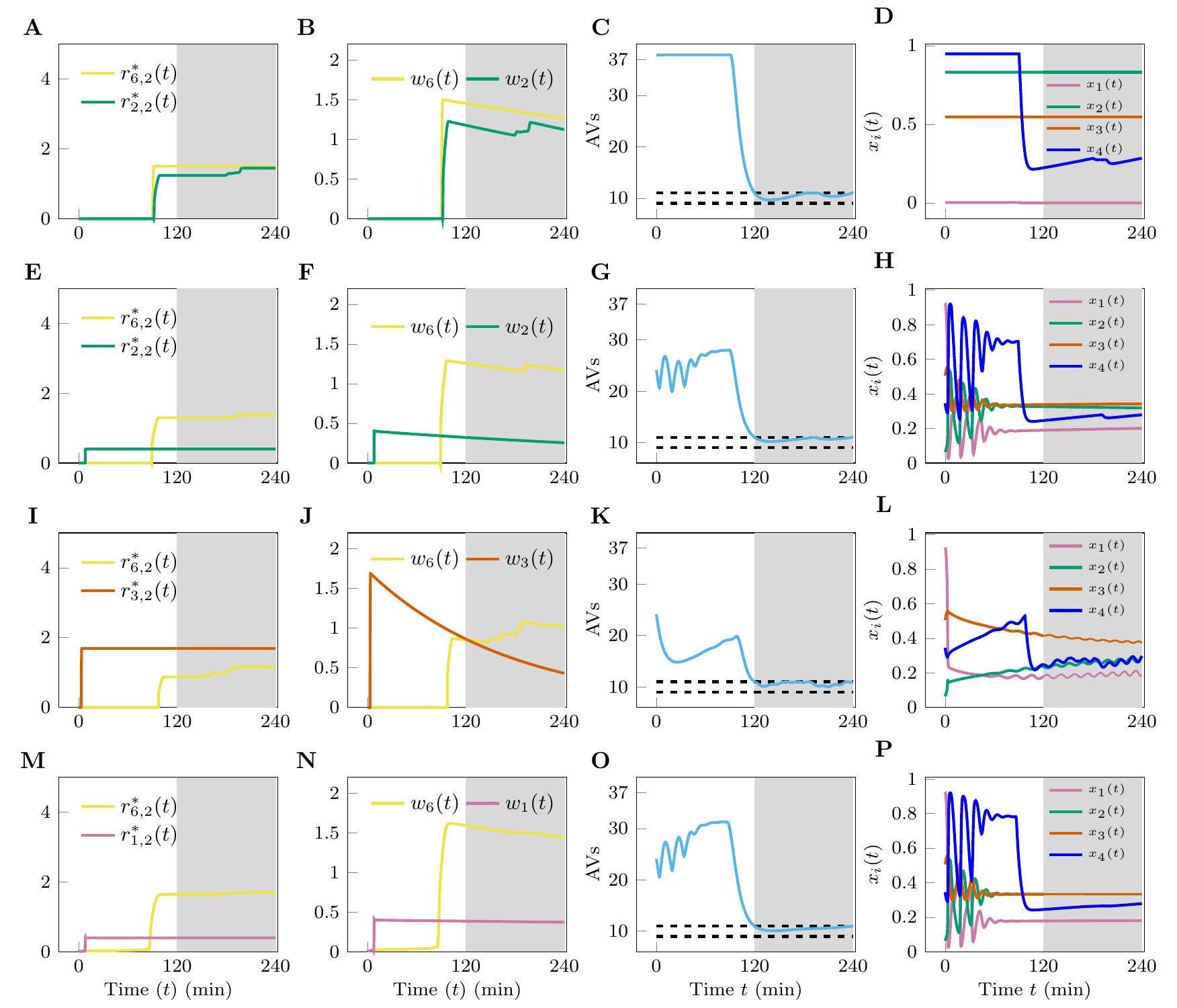}
	\caption{Optimal dual therapies.  
		(\emph{A}--\emph{D}) Panels  \emph{A}--\emph{D} are from a numerical experiment for which we set $C_{\text{Nu}} = C_{\text{En}}= 0.1$ and attempt to use a combination of drugs 2 and 6. (\emph{E}--\emph{H}) Panels  \emph{E}--\emph{H} are from a numerical experiment in which we set $C_{\text{Nu}} = C_{\text{En}}= 0.6$ and attempt to use a combination of drugs 2 and 6. (\emph{I}--\emph{L}) Panels  \emph{I}--\emph{L} are from a numerical experiment in which we set $C_{\text{Nu}} = C_{\text{En}}= 0.6$ and attempt to use a combination of drugs 3 and 6. (\emph{M}--\emph{P}) Panels  \emph{M}--\emph{P} are from a numerical experiment in which we set $C_{\text{Nu}} = C_{\text{En}}= 0.6$ and attempt to use a combination of drugs 2 and 6. The plots on the first column are cumulative drug dosages for the dual therapies considered. The plots on the second column are drug concentrations. The plots in the third column show $x_5(t)$ and the plots in the fourth, rightmost, column show $x_1(t)$, $x_2(t)$, $x_3(t)$, and $x_4(t)$, which we did not attempt to control. In all the simulations, the target value for AV count was 10 (i.e., $x_5^f=10$) and the upper bound on each drug concentration $w_i$ was 2 (i.e., $w_i^{\max} = 2$). The white region corresponds to the time interval $[t_0,t_f]$ when we either upregulate or downregulate the AV count The shaded region corresponds to the time interval $[t_0, t_f]$ when the AV count is maintained within the interval $x_5^f \pm \epsilon$.}
	\label{fig:best_dual}
\end{figure}

\bibliography{auto_ref}
\bibliographystyle{naturemag-doi}

\end{document}


\maketitle

\clearpage
\section*{\LARGE Supplementary Methods}

\section*{Formulation of the Model}

Formulation of Eq.\ (1) was guided by the models of Szyma{\'n}ska et al.\cite{szymanska2015computational} (Ref. 33 in the main text) and Martin et al.\cite{martin2013computational} (Ref. 34 in the main text) mainly as follows. The model of Eq.\ (1) was formulated and parameterized so as to allow the model to predict oscillatory induction of autophagy in response to intermediate drug, energy, and nutrient stress inputs (as illustrated in Figs. 2 and 3), in accord with the predictions of the model of Szyma{\'n}ska et al.\cite{szymanska2015computational}. Moreover, as in both models considered by Martin et al.\cite{martin2013computational}, Eq.\ (1) takes AVs to be turned over constitutively via a pseudo first-order degredative process. Another factor that drove model formulation and parameterization was the availability of measured AV dynamics induced by MTORC1 inhibition\cite{martin2013computational}. Eq.\ (1) was parameterized so as to reproduce the essential aspects of these dynamics (see below for more discussion).

Equation (1) differs from the earlier models of Szyma{\'n}ska et al.\cite{szymanska2015computational} and Martin et al.\cite{martin2013computational} mainly as follows. In the model of Szyma{\'n}ska et al.\cite{szymanska2015computational}, the regulatory influences depicted in Fig. 1 (e.g., mutual inhibition of MTORC1 and ULK1 and negative feedback from ULK1 to AMPK) are not explicitly represented, as is the case in the model of Eq.\ (1), where regulatory influences on enzymatic activities are represented explicitly using Hill functions. Rather, in the model of Szyma{\'n}ska et al.\cite{szymanska2015computational}, regulatory influences emerge from formal representations of the biomolecular interactions considered in the model, which are termed rules\cite{chylek2014}. In other words, Eq.\ (1) provides a model of regulatory influences and their effects, whereas the model of Szyma{\'n}ska et al.\cite{szymanska2015computational} provides a model of biomolecular interactions and their effects, which include emergent regulatory influences. The rules of the model of Szyma{\'n}ska et al.\cite{szymanska2015computational} can be processed automatically by the BioNetGen software package\cite{faeder2009} to obtain a system of 173 coupled ordinary differential equations (ODEs). These equations account for various complexes (e.g., a complex of AMPK and ULK1 that is generated when AMPK docks to a particular site in ULK1) and protein phosphoforms. In contrast, the model of Eq.\ (1) does not track these details. Rather, it simply tracks the activities of AMPK, MTORC1, and ULK1 (and also the activity of VPS34, which was not considered by Szyma{\'n}ska et al.\cite{szymanska2015computational}). In the model of Szyma{\'n}ska et al.\cite{szymanska2015computational}, AMPK, MTORC1, and ULK1 each has numerous states. In contrast, in the model of Eq.\ (1), these protein states are reduced to just two for each protein: active or inactive. 

Although the model of Szyma{\'n}ska et al.\cite{szymanska2015computational} provides a mechanistically detailed representation of biomolecular interactions, it does not include a representation of autophagic vesicle (AV) population dynamics. To include a representation of AV population dynamics in Eq.\ (1), we started with the simple representation of AV production and clearance used in the AV population dynamics model of Martin et al.\cite{martin2013computational}: 
	\begin{equation*}
		\frac{dV}{dt}= P^{\ast} - cV,
	\end{equation*}
where $V$ is cellular AV count, $P^{\ast}$ is a condition-dependent zero-order rate constant for AV production, and $c$ is a pseudo first-order rate constant for clearance of AVs. In our model, we modified this equation by allowing the production rate to be time dependent. In Eq.\ (1) the rate of AV production is a linear function of VPS34 activity, $x_4(t)$. In other words, the rate of AV production is given by $k_3x_4(t)$ (vs. a constant, $P^{\ast}$).

Parameter settings are summarized in Supplementary Tables S1 and S2. These settings are not uniquely determined by data; they were guided by the considerations explained below.

Parameter settings for parameters in the $h$ and $H$ Hill functions were determined first, as follows. For each Hill function, we initially set $r_b = 0$, $r_m = 1$, $\theta = 0.5$, and $n = 2$. (We omit indices in referring to these parameters for convenience.) We then varied parameter values (by hand tuning) to obtain qualitative behavior consistent with that predicted by the model of Szyma{\'n}ska et al.\cite{szymanska2015computational}. The behaviors of the two models are compared directly in Supplementary Fig.\ \ref{fig:model_vs_model_comparison}. In panels \emph{A} and \emph{B} of Supplementary Fig.\ \ref{fig:model_vs_model_comparison}, AV count ($x_5$) and ULK1 activity ($x_2$) are shown, respectively, as a function of time. Initially, in these plots, we consider a nutrient/energy replete condition ($C_{\rm En}=C_{\rm Nu}=1$) without rapamycin (or any other drug). A low dose of rapamycin is added at time $t=100$ min and then a high dose of rapamycin is added at time $t=200$ min. As can be seen, $x_5$ (Supplementary Fig.\ \ref{fig:model_vs_model_comparison}\emph{A}) and $x_2$ (Supplementary Fig.\ \ref{fig:model_vs_model_comparison}\emph{B}) initially have steady low values. After the initial introduction of rapamycin, these quantities begin to oscillate. After the second addition of rapamycin, the two quantities have steady high values. This behavior is qualitatively the same as the behavior predicted by the model of Szyma{\'n}ska et al.\cite{szymanska2015computational} (Supplementary Fig.\ \ref{fig:model_vs_model_comparison}\emph{C}). It should be noted that the study of Szyma{\'n}ska et al.\cite{szymanska2015computational} did not establish that the AMPK-MTORC1-ULK1 network actually exhibits oscillatory behavior; this study only showed that oscillatory behavior is a possible consequence of known regulatory mechanisms. By requiring Eq.\ (1) to reproduce the qualitative nonlinear dynamics of the model of Szyma{\'n}ska et al.\cite{szymanska2015computational}, we made the optimal control problem considered here more of a challenging test of our methodology.

Next, parameter settings for the rate constants $k_1$, $k_2$, $k_3$ and $k_4$ were determined (again through hand tuning). In the study of Martin et al.\cite{martin2013computational}, AV population dynamics were monitored after cells in a nutrient/energy replete condition were treated with a dose of rapamycin or AZD8055 (a catalytic MTOR inhibitor) sufficient to fully inhibit MTORC1 activity. We selected values for the rate constants that allow the model of Eq.\ (1) to roughly reproduce the observed dynamics induced by MTORC1 inhibition in the study of Martin et al.\cite{martin2013computational}. The behaviors predicted by Eq.\ (1) and the model of Martin et al.\cite{martin2013computational} are directly compared in panels \emph{D} and \emph{E} of Supplementary Fig.\ \ref{fig:model_vs_model_comparison}. The AV population dynamics model of Martin et al.\cite{martin2013computational} can be written as follows: $dV/dt=(1+k\delta)P-cV$, where $\delta=0$ indicates a 0 dose of MTORC1 inhibitor, $\delta=1$ indicates a saturating dose of MTORC1 inhibitor, $P$ is the baseline rate of AV production, and $(1+k)P$ is the induced rate of AV production stimulated by a saturating dose of MTORC1 inhibitor. By varying $\delta$ from 0 to 1, we obtain the plots shown in Supplementary Fig.\ \ref{fig:model_vs_model_comparison}\emph{E}. Note that AV dynamics at intermediate values for $\delta$ are not oscillatory, as we would expect from the analysis of Szyma{\'n}ska et al.\cite{szymanska2015computational}.  In contrast, Eq.\ (1) does predict oscillatory AV dynamics at intermediate doses of MTORC1 inhibitor (Supplementary Fig.\ \ref{fig:model_vs_model_comparison}\emph{D}). Importantly, as desired, Eq.\ (1) makes predictions that are in qualitative agreement with the model of Martin et al.\cite{martin2013computational}, in that both models predict that AV dynamics stimulated by MTORC1 inhibitor treatment unfold on a similar timescale and that the maximal range of regulation is similar. In Supplementary Fig.\ \ref{fig:model_vs_data_comparison}, we directly compare the AV dynamics predicted by Eq.\ (1) with AV dynamics measured by Martin et al.\cite{martin2013computational}. As can be seen, Eq.\ (1) is roughly consistent with the data.

Finally, parameter settings for the drug clearance rate constants in Eq.\ (1) ($\delta_1,\ldots,\delta_6$) were set in accordance with measured drug lifetimes reported in the literature, which have half-lives ranging from approximately 1 to 40 h. See Supplementary Table S2 and references cited therein. With this approach, the different drugs considered have different pharmacokinetics, arguably making the optimal control problem more realistic.

\section*{Pseudo-Spectral Optimal Control}\label{sec:PSOC}
	We present here a brief overview of the theory of pseudo-spectral optimal control (PSOC).
	Before discussing the PSOC framework, we briefly review optimal control as well as the difficulties that arise when attempting to solve a general optimal control problem (OCP) analytically.
	Afterwards, we describe how PSOC discretizes the OCP, approximating the original OCP as a nonlinear programming (NLP) problem.
	Approximating the original problem as an NLP is beneficial because there exists a vast literature and many pieces of software capable of solving large-scale NLPs efficiently.
	Finally, we discuss our choices of software, all of which are open-source, and briefly discuss the algorithms they implement.
	\subsection*{Optimal Control}
	The field of optimal control combines aspects of dynamical systems, optimization, and calculus of variations \cite{kirk2012optimal}.
	In words, an optimal control problem is solved by finding a time varying control input $\textbf{u}(t)$ that minimizes a quantity $J(\textbf{x},\textbf{u},t)$ subject to a system's dynamics and other constraints.
	\subsubsection*{General Problem}
	Define the states of the system as $\textbf{x}(t) \in \mathbb{R}^n$, the control inputs as $\textbf{u}(t) \in \mathbb{R}^m$, and time $t \in [t_0,t_f]$ where $t_0 < t_f$.
	The typical form of an optimal control problem for a continuous-time system can be written as,
	\begin{equation}\label{eq:OCP}
	\begin{aligned}
	\min_{\textbf{u}(t)} && &J(\textbf{x}(t),\textbf{u}(t),t) = E\left(\textbf{x}(t_0),\textbf{x}(t_f), t_0, t_f\right) + \int_{t_0}^{t_f} F\left( \textbf{x}(t), \textbf{u}(t), t\right) d t\\
	\text{s.t.} && &\dot{\textbf{x}}(t) = \textbf{f} ( \textbf{x}(t), \textbf{u}(t), t)\\
	&& &\textbf{e}^L \leq \textbf{e}(\textbf{x}(t_0), \textbf{x}(t_f), t_0, t_f) \leq \textbf{e}^U\\
	&& &\textbf{h}^L \leq \textbf{h}(\textbf{x}(t), \textbf{u}(t), t) \leq \textbf{h}^U\\
	&& &t \in [t_0,t_f]
	\end{aligned}
	\end{equation}
	The objective function (or cost function) $J(\textbf{x},\textbf{u},t)$ is composed of two parts, (i) $E : \mathbb{R}^n \times \mathbb{R}^n \times \mathbb{R} \times \mathbb{R} \mapsto \mathbb{R}$ which is a cost associated with the endpoint behavior of the system $\textbf{x}(t_0)$ and $\textbf{x}(t_f)$, and (ii) $F : \mathbb{R}^n \times \mathbb{R}^m \times \mathbb{R} \mapsto \mathbb{R}$ which is a running cost over the entire time interval $[t_0,t_f]$.
	The system dynamics is described by the function $\textbf{f} : \mathbb{R}^n \times \mathbb{R}^m \times \mathbb{R} \mapsto \mathbb{R}^n$.
	Constraints on the endpoints ($\textbf{x}(t_0)$ and/or $\textbf{x}(t_f)$) are described by $\textbf{e} : \mathbb{R}^n \times \mathbb{R}^n \times \mathbb{R} \times \mathbb{R} \mapsto \mathbb{R}^e$.
	While we only specify initial conditions, more complicated relations between the endpoints of the states can be specified as well.
	Finally, path constraints, such as bounds on the states or control inputs, are described by $\textbf{h} : \mathbb{R}^n \times \mathbb{R}^n \times \mathbb{R} \mapsto \mathbb{R}^h$.
	\subsubsection*{Notation for Therapies}
	Let $\mathcal{D} = \{1,2,3,4,5,6\}$ denote the possible drugs we may use (described in the main text) and $\mathcal{T}_k \subseteq \mathcal{D}$ denote the drugs chosen for our therapy such that $|\mathcal{T}_k| = k$.
	Let $\textbf{w}(t) \in \mathbb{R}^k$ denote the drug concentrations and $\textbf{u}(t) \in \mathbb{R}^k$ denote the drug injection rates for \emph{only those drugs chosen to be in the therapy}.
	For example, if we consider the \dual $\mathcal{T}_2 = \{3,6\}$, then
	\begin{equation}
	\begin{aligned}
	\textbf{w}(t) = \left[ \begin{array}{c}
	w_3(t) \\ w_6(t)
	\end{array} \right], && \textbf{u}(t) = \left[ \begin{array}{c}
	u_3(t) \\ u_6(t)
	\end{array} \right]
	\end{aligned}
	\end{equation}
	Those drugs not chosen to be in $\mathcal{T}_k$ are denoted $\mathcal{D} \backslash \mathcal{T}_k$.
	In the example where $\mathcal{T}_k = \{3,6\}$, those drugs not used are $\mathcal{D} \backslash \mathcal{T}_k = \{1,2,4,5\}$.
	If a drug $i \in \mathcal{D} \backslash \mathcal{T}$ then we set $w_i(t) = 0$ for all time $t$.
	
	The drug concentrations appear in the dynamical equations as inhibitory Hill functions $H(w_i(t))$.
	\begin{equation}\label{eq:Hill}
	H(w_i(t)) = r_{m,i} - (r_{m,i} -r_{b,i}) \frac{w_i^{n_i}(t)}{w_i^{n_i}(t)+\theta^{n_i}}
	\end{equation}
	Note that if $i \notin \mathcal{T}_k$, then, as stated previously, $w_i(t) = 0$, and so, by Eq.\ \eqref{eq:Hill}, $H(w_i(t)) = 1$ for all time $t$.
	
	\subsubsection*{The Minimum Drug OCP}
	In the main text, we present a \emph{multi-phase optimal control problem}, i.e., two optimal control problems linked together by enforcing continuity at their interface.
	Despite this added complexity, we can develop a set of necessary conditions for each phase individually and so for now we focus on the single phase problem.
	We will return to the multi-phase problem in the next section that covers the discretization procedure.
	
	Either phase of the OCP presented in the main text can be mapped to the general formulation presented in Eq.\ \eqref{eq:OCP} with the following definitions.
	\begin{itemize}
		\item The state variables $\textbf{x}(t) = \left[\begin{array}{cccccc} x_1(t) & x_2(t) & x_3(t) & x_4(t) & x_5(t) & \textbf{w}^T(t)\end{array} \right]^T \in \mathbb{R}^{5+k}$ and the control input $\textbf{u}(t)\in \mathbb{R}^k$ so that $n = 5+k$ and $m = k$.
		\item The cost function $J = \int_{t_0}^{t_f} u_i(t) d t$ (see Eq.\ (1) in the main text) so that, from Eq.\ \eqref{eq:OCP}, $E \equiv 0$ and $F = \sum_{i \in \mathcal{T}} u_i(t)$.
		\item The system dynamics, as presented in Eq.\ (1), are rewritten here,
		\begin{equation}
		\begin{aligned}
		\dot{\textbf{x}}(t) &= \left[ \begin{array}{c}
		\dot{x}_1(t)\\
		\dot{x}_2(t)\\
		\dot{x}_3(t)\\
		\dot{x}_4(t)\\
		\dot{x}_5(t)\\
		\dot{\textbf{w}}(t)
		\end{array} \right] = \textbf{f}(\textbf{x}(t),\textbf{u}(t)) = \bar{\textbf{f}}(\textbf{x}(t)) + B \textbf{u}(t)\\
		&= \left[ \begin{array}{c}
		(1-x_1)C_\text{Nu} H(w_1) H(w_2) - x_1 h_{12}(x_2)h_{13}(x_3)\\
		(1-x_2) h_{23}(x_3) H(w_3) - x_2 h_{21}(x_1)\\
		(1-x_3) k_1 H(w_4) - C_\text{En}x_2 x_3 H(w_5)\\
		(1-x_4)h_{42}(x_2) H(w_2)H(w_6) - k_2x_4\\
		k_3x_4 - k_4 x_5\\
		- \Delta \textbf{w}(t)
		\end{array} \right] + \left[ \begin{array}{c}
		\boldsymbol{0}_k^T\\
		\boldsymbol{0}_k^T\\
		\boldsymbol{0}_k^T\\
		\boldsymbol{0}_k^T\\
		\boldsymbol{0}_k^T\\
		I_k
		\end{array} \right] \textbf{u}(t)
		\end{aligned}
		\end{equation}
		where $\boldsymbol{0}_k$ is a vector of all zeros of length $k$, $I_k$ is the identity matrix of order $k$, and $\Delta$ is a diagonal matrix with the corresponding rates $\delta_i$ on the diagonal if $i \in \mathcal{T}$.
		For example, if $\mathcal{T} = \{3,6\}$, then 
		\begin{equation}
		\Delta = \left[ \begin{array}{cc}
		\delta_3 & 0 \\ 0 & \delta_6
		\end{array} \right]
		\end{equation}
		Also, note that if $i \notin \mathcal{T}$, then $w_i(t) \equiv 0$ and $H(w_i(t)) = 1$.
		\item The only endpoint constraints are set at the initial time,
		\begin{equation}
		\begin{aligned}
		\textbf{e}(\textbf{x}(t_0),\textbf{x}(t_f),t_0,t_f) = \left[ \begin{array}{c}
		x_1(0) \\ x_2(0) \\ x_3(0) \\ x_4(0) \\ x_5(0) \\ \textbf{w}(0)
		\end{array} \right], && \textbf{e}^L = \textbf{e}^U = \left[ \begin{array}{c}
		x_{1,0} \\ x_{2,0} \\ x_{3,0} \\ x_{4,0} \\ x_{5,0} \\ \textbf{0}_k
		\end{array} \right]
		\end{aligned}  
		\end{equation}
		where $x_{i,0}$ is chosen to either be the steady state value of the system in the absence of control inputs or the time-average of the time evolution of the system if the dynamics, in the absence of control inputs, is oscillatory.
		We assume there is no drug present initially so $w_i(0) = 0$, $i \in \mathcal{D}$.
		\item Finally, the path constraints consist of upper bounds on the drug concentrations and possibly a lower and/or upper bound on the AVs.
		\begin{equation}\label{eq:h}
		\begin{aligned}
		\textbf{h}(\textbf{x}(t),\textbf{u}(t),t) = \left[ \begin{array}{c}
		x_5(t) \\ \textbf{w}(t) \\ \textbf{u}(t)
		\end{array} \right], && \textbf{h}^L = \left[ \begin{array}{c}
		x_5^L \\
		\boldsymbol{0}_k \\
		\boldsymbol{0}_k 
		\end{array} \right], &&
		\textbf{h}^U = \left[ \begin{array}{c}
		x_5^U \\ w^{\max} \boldsymbol{1}_k \\ \infty
		\end{array} \right]
		\end{aligned}
		\end{equation}
		where, for the first phase, $x_5^L = 0$ and $x_5^U = \infty$ but for the second phase we choose $x_5^L = x_5^f - \epsilon$ and $x_5^U + \epsilon$.
		Also, the upper bound on the drug concentration is chosen to be identical for all drugs in the therapy.
	\end{itemize}
	Solving Eq.\ \eqref{eq:OCP} is not a trivial task, and typically there exists no closed form solution.
	Instead one typically must turn to numerical methods, such as PSOC, which we will discuss in the subsequent subsections in some detail.
	Nonetheless, one can derive a set of necessary conditions that any solution to Eq.\ \eqref{eq:OCP} must satisfy using Pontryagin's minimum principle \cite{kirk2012optimal}.
	Developing these types of necessary conditions allows us to construct a set of validation criteria with which we may test the quality of any solution returned by our numerical methods.
	
	A full derivation of Pontryagin's minimum principle is beyond the scope of this work but it is readily available in many standard texts \cite{kirk2012optimal}.
	Here, we present the main results surrounding the Hamiltonian constructed from Eq.\ \eqref{eq:OCP}.
	
	\subsubsection*{Minimizing the Hamiltonian}
	Define a vector of time-varying costates (or adjoint variables) as $\boldsymbol{\lambda}(t) = \left[ \begin{array}{cc} \boldsymbol{\lambda}_{\textbf{x}}^T(t) & \boldsymbol{\lambda}_{\textbf{w}}^T(t) \end{array} \right]^T \in \mathbb{R}^{5+k}$ so that $\boldsymbol{\lambda}_{\textbf{x}}(t) \in \mathbb{R}^5$ and $\boldsymbol{\lambda}_{\textbf{w}}(t) \in \mathbb{R}^k$.
	The Hamiltonian of the OCP in Eq.\ \eqref{eq:OCP} is defined as,
	\begin{equation}\label{eq:H}
	\begin{aligned}
	H(\boldsymbol{\lambda}, \textbf{x}, \textbf{u}, t) &= F(\textbf{x}, \textbf{u}, t) + \boldsymbol{\lambda}^T \textbf{f}(\textbf{x},\textbf{u},t)\\
	&=\sum_{i \in \mathcal{T}} u_i + \boldsymbol{\lambda}^T \bar{\textbf{f}}(\textbf{x}) + \boldsymbol{\lambda} B \textbf{u}
	\end{aligned}
	\end{equation}
	where $\boldsymbol{\lambda}(t) \in \mathbb{R}^n$ are the costates (or adjoint variables).
	A solution to Eq.\ \eqref{eq:OCP} must also be a solution of the following minimization problem.
	\begin{equation}\label{eq:HMC}
	\begin{aligned}
	\min_{\textbf{u}(t)} && &H(\boldsymbol{\lambda},\textbf{x},\textbf{u},t)\\
	\text{s.t.} && &\textbf{h}^L \leq \textbf{h}(\textbf{x},\textbf{u},t) \leq \textbf{h}^U
	\end{aligned}
	\end{equation}
	To solve Eq.\ \eqref{eq:HMC}, we define the associated Lagrangian,
	\begin{equation}\label{eq:Hbar}
	\begin{aligned}
	\bar{H} (\boldsymbol{\mu},\boldsymbol{\lambda}, \textbf{x}, \textbf{u}, t) &= H(\boldsymbol{\lambda}, \textbf{x}, \textbf{u}, t) + \boldsymbol{\mu}^T \textbf{h}(\textbf{x},\textbf{u},t)\\
	&= \sum_{i \in \mathcal{T}} u_i + \boldsymbol{\lambda}^T \bar{\textbf{f}}(\textbf{x}) + \boldsymbol{\lambda}^T B \textbf{u} + \mu_{x_5} x_5 + \boldsymbol{\mu}_{\textbf{w}}^T \textbf{w} + \boldsymbol{\mu}_{\textbf{u}}^T \textbf{u}
	\end{aligned}
	\end{equation}
	where $\boldsymbol{\mu} = \left[ \begin{array}{ccc} \mu_{x_5} & \boldsymbol{\mu}_{\textbf{w}}^T & \boldsymbol{\mu}_{\textbf{u}}^T \end{array} \right]^T \in \mathbb{R}^h$ is the copath vector with components associated with the components of the vector of path constraints in \eqref{eq:h}.
	A solution to Eq.\ \eqref{eq:HMC}, and thus to our original OCP, must satisfy,
	\begin{equation}\label{eq:dHdu}
	\begin{aligned}
	\frac{\partial \bar{H}}{\partial \textbf{u}} = \boldsymbol{1}_k + B^T \boldsymbol{\lambda} + \boldsymbol{\mu}_{\textbf{u}} = \boldsymbol{0}
	\end{aligned}
	\end{equation}
	where the costates evolve according to the dynamical equation,
	\begin{equation}
	\dot{\boldsymbol{\lambda}} = -\frac{\partial \bar{H}}{\partial \textbf{x}} = - \left( \frac{\partial \bar{\textbf{f}}}{\partial \textbf{x}} \right)^T \boldsymbol{\lambda} + \left[ \begin{array}{c}
	\boldsymbol{0}_4 \\ \mu_{x_5} \\ \boldsymbol{\mu}_{\textbf{w}}
	\end{array} \right]
	\end{equation}
	The optimal control input $u_i(t)$, $i \in \mathcal{T}$, must satisfy the complementarity condition \cite{nocedal2006numerical, ross2015primer} 
	\begin{equation}\label{eq:complement}
	\left\{ \begin{aligned}
	u_i(t) = 0 && \text{if} && \mu_i(t) < 0\\
	u_i(t) \geq 0 && \text{if} && \mu_i(t) = 0\\
	u_i(t) \rightarrow \infty  && \text{if} && \mu_i(t) > 0
	\end{aligned} \right.
	\end{equation}
	Combining Eqs.\ \eqref{eq:dHdu} and \eqref{eq:complement}, we can relate $\boldsymbol{\mu}_{\textbf{u}}$ to the time-varying costates by noting from the structure of $B$, $B^T \boldsymbol{\lambda} = \boldsymbol{\lambda}_{\textbf{w}}$ so that,
	\begin{equation}
	\boldsymbol{\mu}_{\textbf{u}}(t) = -\boldsymbol{1}_k - \boldsymbol{\lambda}_{\textbf{w}}(t)
	\end{equation}
	Thus, if $\lambda_{w_i} > -1$ then $u_i = 0$, but if $\lambda_{w_i} = -1$, then all we can say is that $u_i \geq 0$. 
	When $\lambda_{w_i} > -1$, the optimal control is said to have a \emph{singular arc} (see chapter 5 in \cite{kirk2012optimal}).
	Despite the technical difficulties, we have arrived at our first set of validation conditions, that is,
	\begin{equation}
	\begin{aligned}
	u_i \cdot (\lambda_{w_i}-1) = 0, && \forall i \in \mathcal{T}
	\end{aligned}
	\end{equation}
	Let us now assume that we have solved Eq.\ \eqref{eq:HMC}, that is,
	\begin{equation}
	\begin{aligned}
	\mathcal{H}(t) = \min_{\textbf{u} \in \mathbb{U}} H(\boldsymbol{\lambda},\textbf{x},\textbf{u},t)
	\end{aligned}
	\end{equation}
	where $\mathbb{U}$ is the set of feasible control inputs, i.e., they satisfy all of the constraints imposed by Eq.\ \eqref{eq:OCP}.
	The evolution of the Hamiltonian at the optimal solution can be written,
	\begin{equation}
	\frac{d \mathcal{H}}{d t} = \frac{\partial H}{\partial t}
	\end{equation}
	where, since in our OCP, $H$ does not explicitly depend on time, we expect that $d\mathcal{H} / d t = 0$ and so $\mathcal{H}$ should be constant.
	This is the second validation condition.
	
	While in the paper and the supplementary sections we display time traces of the states and the control inputs as they are the quantities of interest to the general reader, we are also able to access the costate and copath time traces, as well as the time trace of the Hamiltonian.
	In Fig. \ref{fig:VV} we show a typical set of output that we use for measuring the quality of our returned numerical solution.
	The sample shows a \mono where $\mathcal{T} = \{4\}$.
	Panel (a) shows the level of AVs, $x_5(t)$, and panel (b) shows the drug concentration $w_4(t)$.
	Panel (c) contains the copath associated with the level of AVs, $\mu_{x_5}(t)$.
	Note that during the first phase when there is no finite bound on $x_5(t)$ the copath $\mu_{x_5}(t) = 0$, while during second phase if $\mu_{x_5}(t) \neq 0$ then $x_5(t) = x_5^f \pm \epsilon$.
	In panel (d) we plot the other copath $\mu_{w_4}(t)$.
	The control input $u_4(t)$ itself is shown in panel (e) along with the costate $\lambda_{w_4}(t)$ in panel (f).
	Note that the times at which $u_4(t) > 0$ correspond to times when $\lambda_{w_4}(t) = -1$ as expected.
	Panel (g) plots the time evolution of the Hamiltonian evaluated at the optimal solution.
	Note that the $y$-axis is scaled by $10^{-2}$.
	We see that $\mathcal{H} \approx const$ within each phase, with a jump occurring at the interface between the two phases.
	As we cannot say anything about the value of the Hamiltonian at the interface, a discontinuity at this point in time can be expected.
	
	\subsection*{Discretization of the OCP}
	As presented in the previous subsection, we have seen that the set of necessary conditions which must be satisfied consist of a system of coupled nonlinear differential equations for $\textbf{x}(t)$ and $\boldsymbol{\lambda}(t)$ along with a set of non-trivial constraints.
	Searching for an analytic solution is unlikely to be successful and so instead we turn to pseudospectral optimal control (PSOC).
	
	In short, PSOC is a methodology by which one may discretize an OCP, approximating the integrals by quadratures and the time-varying states and control inputs with interpolating polynomials.
	
	The key to PSOC is choosing the discretization points properly.
	Let $\{\tau_i\}$, $i = 0, \ldots, N$, denote the discretization points.
	Typically these are chosen as the roots of an orthogonal polynomial such as a Legendre polynomial or a Chebyshev polynomial of order $N$.
	For some popular choices of discretization schemes see \cite{rao2009survey}.
	For concreteness, we will assume that $\tau_0 = -1$ and $\tau_N = 1$, i.e., we are using a discretization scheme that includes the endpoints and is normalized by the mapping,
	\begin{equation}
	t = \frac{t_f-t_0}{2} \tau + \frac{t_f+t_0}{2}
	\end{equation}
	For the discretization scheme chosen, we also compute the associated quadrature weights.
	For instance, if we choose the roots of a Legendre polynomial as the discretization scheme, the associated quadrature weights can be found in the typical way for Gauss quadrature.
	The time-varying states and control inputs are found by approximating them with a Lagrange interpolating polynomial.
	\begin{equation}\label{eq:dxu}
	\begin{aligned}
	\textbf{x}(\tau) &\approx \hat{\textbf{x}}(\tau) = \sum_{i=0}^N \hat{\textbf{x}}_i L_i(\tau)\\
	\textbf{u}(\tau) &\approx \hat{\textbf{u}}(\tau) = \sum_{i=0}^N \hat{\textbf{u}}_i L_i(\tau)
	\end{aligned}
	\end{equation}
	The Lagrange interpolating polynomials are defined as,
	\begin{equation}
	L_i(\tau) = \prod_{j=0,j\neq i}^N \frac{\tau - \tau_j}{\tau_i - \tau_j}
	\end{equation}
	Note that the Lagrange interpolating polynomials satisfy the isolation property, that is, $L_i(\tau_j) = \delta_{i,j}$.
	We can thus construct a set of algebraic equations corresponding to the discretization points $\{\tau_i\}$.
	Define $D_{k,i} = \frac{d L_i}{d\tau}(\tau_k)$ so that the derivative of the states at the discretization points can be approximated as,
	\begin{equation}\label{eq:dxhat}
	\dot{\hat{\textbf{x}}}(\tau_k) = \sum_{i=0}^N \hat{\textbf{x}}_i D_{k,i}
	\end{equation}
	With Eqs.\ \eqref{eq:dxu} and \eqref{eq:dxhat}, we can approximate the original system of $n$ differential equations as $n(N+1)$ algebraic equations.
	\begin{equation}\label{eq:discstate}
	\begin{aligned}
	\sum_{i=0}^N D_{k,i} \hat{\textbf{x}}_i - \frac{t_f-t_0}{2} \textbf{f}(\hat{\textbf{x}}_k, \hat{\textbf{u}}_k, \tau_k) = \boldsymbol{0}_n, && k = 1,\ldots,N\\
	\hat{\textbf{x}}_N - \hat{\textbf{x}}_0 - \sum_{k=1}^N \sum_{i=0}^N w_k D_{k,i} \hat{\textbf{x}}_i = \boldsymbol{0}_n
	\end{aligned}
	\end{equation}
	The last set of algebraic constraints arise from the consistency condition $\int_{t_0}^{t_f} \dot{\textbf{x}}(t) d t = \textbf{x}(t_f) - \textbf{x}(t_0)$.
	Similarly to the consistency condition, the integral in the cost function is approximated as,
	\begin{equation}
	J = \int_{t_0}^{t_f} F(\textbf{x},\textbf{u},t) \approx \hat{J} = \frac{t_f-t_0}{2} \sum_{k=1}^N F(\hat{\textbf{x}}_k, \hat{\textbf{u}}_k, \tau_k)
	\end{equation}

	The discretized approximation of the original OCP is compiled into the following nonlinear programming (NLP) problem.
	\begin{equation}\label{eq:NLP}
	\begin{aligned}
	\min_{\textbf{u}_i} && &\hat{J} = \frac{t_f-t_0}{2} \sum_{k=1}^N F(\hat{\textbf{x}}_k, \hat{\textbf{u}}_k, \tau_k)\\
	\text{s.t.} && &\sum_{i=0}^N D_{k,i} \hat{\textbf{x}}_i - \frac{t_f-t_0}{2} \textbf{f}(\hat{\textbf{x}}_k,\hat{\textbf{u}}_k, \tau_k) = \boldsymbol{0},\ k = 1,\ldots,N\\
	&& &\hat{\textbf{x}}_N - \hat{\textbf{x}}_0 - \sum_{k=1}^N \sum_{i=0}^N w_k D_{k,i} \hat{\textbf{x}}_i = \boldsymbol{0}\\
	&& &\textbf{e}^L \leq \textbf{e}(\hat{\textbf{x}}_0,\hat{\textbf{x}}_N, \tau_0,\tau_N) \leq \textbf{e}^U\\
	&& &\textbf{h}^L \leq \textbf{h}(\hat{\textbf{x}}_k, \hat{\textbf{u}}_k, \tau_k) \leq \textbf{h}^U
	\end{aligned}
	\end{equation}

	With the above results, we now present the application to the full multi-phase optimal control problem.
	In general, let us assume there are $p$ phases where $p=2$ in our problem.
	Each phase is active within the interval $t \in [t_0^{(p)},t_f^{(p)}]$.
	In each phase there is a cost function $J^{(p)}$, a dynamical system $\textbf{f}^{(p)}$, a set of endpoint constraints $\textbf{e}^{(p)}$, and a set of path constraints $\textbf{h}^{(p)}$.
	If two phases, $p$ and $q$, are linked, then there also exists a set of linkage constraints $\Phi^{(p,q)}$.
	\begin{equation}\label{eq:mpOCP}
	\begin{aligned}
	\min_{\textbf{u}^{(p)}} && &\sum_{p = 1}^P J^{(p)} = \sum_{p=1}^P \int_{t_0^{(p)}}^{t_f^{(p)}} F^{(p)}(\textbf{x}^{(p)},\textbf{u}^{(p)},t) d t\\
	\text{s.t.} && &\dot{\textbf{x}}^{(p)}(t) = \textbf{f}^{(p)}(\textbf{x}^{(p)}, \textbf{u}^{(p)}, t)\\
	&& &\textbf{h}^{L,(p)} \leq \textbf{h}^{(p)}(\textbf{x}^{(p)},\textbf{u}^{(p)},t) \leq \textbf{h}^{U,(p)}\\
	&& &\textbf{e}^{L,(p)} \leq \textbf{e}^{(p)}(\textbf{x}^{(p)}(t_0^{(p)}), \textbf{x}^{(p)}(t_f^{(p)}), t_0^{(p)}, t_f^{(p)}) \leq \textbf{e}^{U,(p)}\\
	&& &\Phi^{L,(p,q)} \leq \Phi^{(p,q)}(\textbf{x}^{(p)},\textbf{x}^{(q)},\textbf{u}^{(p)},\textbf{u}^{(q)}) \leq \Phi^{U,(p,q)}
	\end{aligned}
	\end{equation}
	Each phase is discretized with its own set of points, $\{\tau_i^{(p)}\}$ so that,
	\begin{equation}
	\textbf{x}^{(p)}(\tau) \approx \hat{\textbf{x}}^{(p)}(\tau) = \sum_{i=1}^N \hat{\textbf{x}}_i^{(p)} L_i(\tau)
	\end{equation}
	so that the full multi-phase NLP is,
	\begin{equation}\label{eq:dmpOCP}
	\begin{aligned}
	\min_{\textbf{u}_i^{(p)}} && &\sum_{p=1}^P \frac{t_f^{(p)} - t_0^{(p)}}{2} \sum_{k=1}^N F^{(p)}(\hat{\textbf{x}}_k^{(p)},\hat{\textbf{u}}_k^{(p)},\tau_k)\\
	\text{s.t.} && &\sum_{i=0}^N D_{k,i} \hat{\textbf{x}}_i^{(p)} - \frac{t_f^{(p)}- t_0^{(p)}}{2} \textbf{f}^{(p)} (\hat{\textbf{x}}^{(p)}_k, \hat{\textbf{u}}_k^{(p)},\tau_k) = \boldsymbol{0}_n, \quad p = 1,\ldots,P, \quad k = 1,\ldots,N\\
	&& &\hat{\textbf{x}}_N^{(p)} - \hat{\textbf{x}}_0^{(p)} - \frac{t_f^{(p)}-t_0^{(p)}}{2} \sum_{k=1}^N \sum_{i=0}^N w_k D_{k,i} \hat{\textbf{x}}_i = \boldsymbol{0}_n, \quad p = 1,\ldots,P\\
	&& &\textbf{e}^{L,(p)} \leq \textbf{e}^{(p)}(\hat{\textbf{x}}_0^{(p)},\hat{\textbf{x}}_N^{(p)},t_0^{(p)},t_f^{(p)}) \leq \textbf{e}^{U,(p)}, \quad p = 1,\ldots,P\\
	&& &\textbf{h}^{L,(p)} \leq \textbf{h}^{(p)} (\hat{\textbf{x}}^{(p)}_k, \hat{\textbf{u}}_k^{(p)}, \tau_k) \leq \textbf{h}^{U,(p)}, \quad k = 1, \ldots,N, \quad p = 1,\ldots P\\
	&& &\Phi^{L,(p,q)} \leq \Phi^{(p,q)}(\hat{\textbf{x}}_0^{(p)},\hat{\textbf{u}}^{(p)}_0, \hat{\textbf{x}}_N^{(q)}, \hat{\textbf{u}}_N^{(q)}) \leq \Phi^{U,(p,q)}, \quad p,q = 1,\ldots,P
	\end{aligned}
	\end{equation}
	To perform the discretization described in this subsection, we use the open-source C++ PSOC package $\mathcal{PSOPT}$ \cite{becerra2010solving}.
	
	Next we show that Eq.\ \eqref{eq:dmpOCP} can be expressed in the typical NLP form \cite{nocedal2006numerical}.
	Let $\textbf{z}^{(p)}$ contain all of the variables for phase $p$.
	\begin{equation}
	\textbf{z}^{(p)} = \left[ \begin{array}{c}
	\hat{\textbf{x}}_0^{(p)} \\ \vdots \\ \hat{\textbf{x}}_N^{(p)} \\ \hat{\textbf{u}}_0^{(p)} \\ \vdots \\ \hat{\textbf{u}}_N^{(p)}
	\end{array} \right] \in \mathbb{R}^{(n+m)}
	\end{equation}
	Next, let $\textbf{z}$ contain the variables for every phase,
	\begin{equation}
	\textbf{z} = \left[ \begin{array}{c}
	\textbf{z}^{(1)} \\ \vdots \\ \textbf{z}^{(P)}
	\end{array} \right] \in \mathbb{R}^{(N+1)(n+m)}
	\end{equation}
	With some algebraic manipulation, the entire discretized multi-phase OCP can be rewritten as an NLP in the typical form.
	\begin{equation}\label{eq:NLPs}
	\begin{aligned}
	\min_{\textbf{z}} && &c(\textbf{z})\\
	\text{s.t.} && &\textbf{g}(\textbf{z}) = \boldsymbol{0}\\
	&& &\textbf{d}(\textbf{z}) \leq \boldsymbol{0}
	\end{aligned}
	\end{equation}
	To solve the large-scale NLP in Eq.\ \eqref{eq:NLPs} we employ an interior-point algorithm \cite{nocedal2006numerical}.
	Specific details of the algorithm are outside the scope of this paper. We used the open-source C++ package Ipopt \cite{wachter2006implementation} to solve each instance of Eq.\ \eqref{eq:NLPs}.
	We direct interested readers who would like to learn more about the technical detailed involved when solving Eq.\ \eqref{eq:NLPs} to the documentation provided with Ipopt.

	The optimal solution returned, $\textbf{z}^\ast$, is separated into its component parts; first by splitting it into the phases $\textbf{z}^{(p)\ast}$, and second by reconstructing the discrete states and contrlol inputs, $\hat{\textbf{x}}_i^\ast$ and $\hat{\textbf{u}}_i^\ast$.
	The continuous time control inputs and states are then reconstructed using the Lagrange interpolating polynomials in Eq.\ \eqref{eq:dxu}.
	With the continuous time states and control inputs, $\textbf{x}^\ast(t)$ and $\textbf{u}^\ast(t)$, we then verify that the necessary conditions are met to within an acceptable tolerance.

	\clearpage
	\section*{\LARGE Supplementary Note}

	\section*{The Response of AVs to Constant Perturbation by Dual Therapies}\label{response_si_2drugs}
	
	Before solving the optimal control problem presented in the main text, we explore the capabilities of the \duals in terms of \up and \down with constant drug concentration as we did in Fig.\ 3 of the main manuscript.
	There, we plotted the long-time response of the system to an individual time-constant drug concentration ($w$) perturbation for the two sets of parameters $C_ {\text{Nu}} = C_ {\text{En}} = 0.1$ and $C_{\text{Nu}} = C_{\text{En}} = 0.6$.
	Similarly, in Fig.\ \ref{fig:constinput_2drugs_0101} and \ref{fig:constinput_2drugs_0606}, we plot the long-time system AV response for the case of \duals with time-constant drug concentration perturbations.

	In Fig.\ \ref{fig:constinput_2drugs_0101}, we set the parameters $C_ {\text{Nu}} = C_{\text{En}}= 0.1$.
	For these parameter values, in the absence of any drugs (control inputs), the sole attractor of the dynamical system corresponds to a high AV count ($\approx 37$).
	Fig.\ \ref{fig:constinput_2drugs_0101} shows the long-time AV response when the system is perturbed by different combinations of constant inputs.
	Note that those subsets that contain either drug $2$ or $6$ are capable of driving the AVs to zero if $w^{\max}$ is made large enough (pairs $\{2,3\}$, $\{2,4\}$, $\{2,6\}$, $\{3,6\}$, $\{4,6\}$, and $\{1,6\}$).
	For each pair $\{i,j\}$, we set $w_i = w_j$ and all other values $w_k = 0$, $k \neq i$ and $k \neq j$.
	The pair $\{3,4\}$ on the other hand is only capable of driving the AVs to $\approx 10$ where any increase of $w^{\max}$ afterwards can produce no further results.
	Also, \dual $\{1,5\}$ is incapable of \down.

	In Fig.\ \ref{fig:constinput_2drugs_0606}, we set the parameters $C_{\text{Nu}} = C_{\text{En}}=0.6$, for which the free evolution of the system is periodic (see Fig. 2 in the main text), and show the same long-time AV response results under constant drug concentration perturbation.
	For all \duals shown, small drug concentrations are unable to remove the oscillations present (denoted by the shaded regions).
	Similar to Fig. \ref{fig:constinput_2drugs_0101}, we see that all drug combinations that contain either drug $2$ or $6$ are capable of driving the level of AVs to zero for $w^{\max}$ set large enough.
	Also, \dual $\{3,4\}$, as before, is only able to reduce the AVs level to $\approx 10$ while the \dual $\{1,5\}$ instead \ups the AVs.
	
	\section*{Exhaustive Analysis of Two-Drug Combinations}\label{sec:twodrug}
	
	In this section, we present simulation results for all possible \duals.
	First, we set both the parameters $C_\text{Nu} = C_\text{En} = 0.1$ for which the number of AVs at steady state in the absence of control inputs is equal to $\approx 37$.
	We attempt to \down the number of AVs using pairs of drugs from the set $\{2, 3, 4, 6\}$ so that there are a total of $\binom{4}{2}= 6$ combinations.
	A pair of drugs drawn from this set is called a \dual.
	If $\{i,j\}$ is a \dual, then we say $\{i\}$ and $\{j\}$ are its component \monos.

	The goal is to investigate our ability to \down the number of AVs from the steady state value $\approx 37$ to a lower value in a specified control time interval $[0,t_0]$ and, subsequently, to maintain the number of AVs near the target level for a second time interval $[t_0,t_f]$, by using each different \dual.
	We say a \dual is \emph{viable} if it is capable of performing the goal stated.
	A \dual is deemed efficient if;
	\begin{itemize}
		\item the \dual is viable while at least one of its component \monos is not, and
		\item the total amount of drugs provided by the \dual is less than either of the component \monos.
	\end{itemize}
	To compare the efficiencies of the \duals we define $r^\ast_{i,k}(t) = \int_0^t u_i^*(\tau) d\tau$ as the total amount of drug $i$ administered at time $t$ as part of a $k = \text{dual}$ or $k = \text{mono}$ and introduce the quantities $\rho_i$ and $\tau_i$. 
	\begin{equation}\label{eq:ratio}
	0 \leq \rho_{i} = \frac{r^\ast_{i,\text{dual}}(t_f)}{r^\ast_{i,\text{mono}} (t_f) }\leq 1,
	\end{equation}
	Note that  $r^\ast_{i,\text{dual}}(t_f) \leq r^\ast_{i,\text{mono}} (t_f)$, as otherwise the solution of the \dual optimal control problem would be suboptimal with respect to the case that only drug $i$ is used.
	We also define the ratio
	\begin{equation}\label{eq:activation}
	\tau_{i} = \frac{\bar{t}_{i,\text{dual}}-\bar{t}_{i,\text{mono}}}{\bar{t}_{i,\text{mono}} }
	\end{equation}
	where $\bar{t}_{i,\text{dual}}$ is the time when drug $i$ is activated (that is, the earliest time at which the drug injection rate is nonzero) as a part of a \dual and $\bar{t}_{i,\text{mono}} $ is the time when drug $i$ is activated as a \mono.
	Note that $\tau_i > 0$ ($\tau_i < 0$) indicates a later (earlier) activation time of drug $i$ as a part of \dual compared to as a \mono.
	
	For our simulations, we set the upper bound of the drug concentrations to $w_i^{\max} = 2$ for each drug $i$, the time at which we apply the upper bound to the AVs to $t_0=120$ minutes, the time at which we end the simulation to $t_f=240$ minutes, and we set the initial condition $\textbf{x}(0)$ to be equal to the steady state solution of the system in the absence of control inputs with parameters $C_{\text{En}} = C_{\text{Nu}} = 0.1$.
	In Fig.\ \ref{fig:Nu01En01}, we plot the total drug administered $r_i(t) = \int_0^t u_i(\tau) d\tau$ in the interval $[0, t_f]$.
	The plots on the diagonal panels, labeled $(u_i , u_i)$, correspond to the \monos and the plots on the upper triangular panels, labeled $(u_i,u_j)$, correspond to the \duals.
	Symmetric to each upper triangular panel $(u_i , u_j)$, the corresponding lower triangular panel $(u_j,u_i)$ contains the values of the ratios $\rho_i$ and $\tau_i$ in Eqs.\ \eqref{eq:ratio} and \eqref{eq:activation}, respectively.
	
	We notice from Fig.\ 3\emph{A} in the main text that the only \monos which can \down the number of AVs from $\approx 37$ to $\approx 10$, with $w_i \leq 2$, is $\{4\}$.
	Thus, the red crosses in panels $(u_2,u_2)$, $(u_3 , u_3)$ and $(u_6,u_6)$ in Fig.\ \ref{fig:Nu01En01} indicate that those \monos cannot solve the \down problem.
	Clearly, \duals $\{2,4\}$, $\{3,4\}$ and $\{4,6\}$ are viable as drug $\{4\}$ as a \mono is viable.
	On the other hand, the \duals $\{2,3\}$ and $\{3,6\}$ are not viable.
	The most interesting \dual is $\{2,6\}$ as neither component \mono is viable yet as a pair they are viable
	Thus by our stated goal and definitions, the \dual $\{2,6\}$ is efficient according to our criteria.
	Also, \dual $\{3, 4\}$ is deemed efficient as the total consumption of drug $4$ is much lower ($\rho_4=0.29$) than the total consumption of drug $4$ as a \mono as shown in panel $(u_3,u_4)$ in Fig.\ \ref{fig:Nu01En01}.
	We also observe the faster response of drug $4$ as a part of the $\{3,4\}$ \dual than its response as a \mono because $\tau_4 = 0.32 > 0$. 
	
	In Fig.\ \ref{fig:Nu01En01_other}, we consider the \duals by combining one of the \down drugs, 2, 3, 4 or 6, with one of the \up drugs, 1 or 5.
	A red cross in a panel again represents a \mono or a \dual that is not viable.
	While the \duals $\{1,4\}$ and $\{4,5\}$ are viable, they are not efficient as neither drugs $1$ nor $5$ are used (non-zero).
	%
	
	In Fig.\ \ref{fig:Nu06En06_down}, we present detailed results when we set the parameters $C_\text{En}=C_\text{Nu} = 0.6$, for which the dynamics in the absence of control inputs is oscillatory.
	In our numerical experiments, we attempt to \down the number of AVs from its initial periodic behavior to $x_5(t_0) \approx 10$ and to maintain the number of AVs near that value for the time interval $[t_0=120,t_f=240]$.
	The red cross in panel $(u_6,u_6)$ indicates the inability of drug $6$ as a \mono to \down the AVs to the desired level.
	However, we found this drug to be particularly beneficial when used as a component in a \dual.
	We find that while all \duals are viable, the most efficient \dual is $\{2,6\}$, as the total amount of drug $2$ required is reduced by more than five folds when compared to the \mono $\{2\}$.
	A comparison with drug $6$ alone is not possible as drug $6$ as a \mono is not viable.
	The \dual $\{3,6\}$ is also efficient by our definition, but only slightly as the amount of drug $3$ used is hardly reduced, $\rho_3=0.96$.
	For all other \duals, one of the component drugs is never activated so while they may by viable, we do not consider them efficient.

	In Fig. \ref{fig:Nu06En06_up}, we summarize the results when we attempt to \up the number of AVs to $\approx 37$ in the same control time interval $[0,t_0]$ and, subsequently, maintain the number of AVs throughout the time interval $[t_0, t_f]$ by using \dual $\{1,5\}$.
	We observe that, while the \dual $\{1,5\}$ is viable, it is not efficient as drug $1$ is never activated and so we must use the same amount of drug $5$ as when it is used as a \mono.
	
	In Fig.\ \ref{fig:Nu06En06_other}, we consider the \duals by combining one of the \down drugs, 2, 3, 4 or 6, with one of the \up drugs, 1 or 5.
	We observe that the \duals $\{1,6\}$ and $\{5,6\}$ are only efficient when $C_{\text{En}} = C_{\text{Nu}} = 0.6$.
	The other \duals while viable are not efficient as the \up component (either $1$ or $5$) is never activated (that is, non-zero).

\clearpage
\renewcommand{\refname}{References Cited in Supplementary Information}

\section*{Supplementary Tables}
\renewcommand{\tablename}{Supplementary Table}
\renewcommand{\thetable}{S\arabic{table}}

\clearpage

\label{table1}
	\begin{table}[tbhp!]
		\centering
		\caption{Parameters of the model (Eq.\ (1)). See ``Formulation of the Model'' in Supplementary Methods for discussion. The parameter values are dimensionless except as indicated.}
		{\setlength\doublerulesep{0.5pt}   
			\begin{tabular}{rlcrl}
				\toprule[1 pt] \midrule
				Parameter  & Value & \quad &  Parameter   & Value  \\
				\midrule \midrule 
				$r_{b,12}$  & 0   & & $k_1$  &  $1.00\times 10^{-1}$  \\
				$r_{m,12}$  & $1.00\times 10^1$  &&  $k_2$   & $3.00\times 10^{-1}$ \\
				$\theta_{12}$  &  $3.00\times 10^{-1}$   &&  $k_3$  & $4.00\times 10^0$\\
				$n_{12}$   & $4.00\times 10^0$  &&  $k_4$   & $1.00\times 10^{-1}$    \\
				$r_{b,13}$  &  0   &&  $\delta_1$   & $3.10\times 10^{-4}$  \\
				$r_{m,13}$  & $1.00\times 10^1$  &&  $\delta_2$  & $1.93\times 10^{-3}$  \\
				$\theta_{13}$  & $6.00\times 10^{-1}$   && $\delta_3$   & $5.78 \times 10^{-3}$ \\
				$n_{13}$  & $6.00\times 10^0$  && $\delta_4$   & $1.15 \times 10^{-2}$   \\
				$r_{b,23}$  & 0   &&  $\delta_5$  & $2.31 \times 10^{-3}$   \\
				$r_{m,23}$  & $6.00\times 10^0$&   &   $\delta_6$   & $1.16 \times 10^{-3}$  \\
				$\theta_{23}$  & $1.00\times 10^0$   &&  $r_{b}$ & 0  \\
				$n_{23}$  & $4.00\times 10^0$   &&   $r_{m}$   & $1.00\times 10^0$   \\
				$r_{b,21}$  & $1.00\times 10^{-1}$   &&  $\theta$   & $5.00\times 10^{-1}$  \\
				$r_{m,21}$  & $6.00\times 10^0$   &&  $n$   & $2.00\times 10^0$   \\
				$\theta_{21}$  & $6.00\times 10^{-1}$   && $T$ &  $1.00\times 10^0$ (min)\\
				$n_{21}$  & $4.00\times 10^0$   && &  \\
				$r_{b,42}$  & $1.00\times 10^{-1}$   && &   \\
				$r_{m,42}$  & $6.00\times 10^0$   &&  & \\
				$\theta_{42}$  & $5.00\times 10^{-1}$   &&  & \\
				$n_{42}$  & $4.00\times 10^0$   && &   \\
				\midrule
				\bottomrule[1pt]
			\end{tabular}
		}
		\label{tab:param}
	\end{table}
	
	\clearpage
	
	\begin{table}[tbhp!]
		\centering
		\caption{Summary of measured drug half-lives used to set values for the drug clearance rate constants $\delta_1,\ldots,\delta_6$ in Eq.\ (1). Each half-life, $t_{1/2,i}$, is the measured half-life of a representative of drug type $i$. See the references cited in the table for details about the drugs and measurements. }
		{\setlength\doublerulesep{0.5pt}   
			\begin{tabular}{cccccccl}
				\toprule[1 pt] \midrule
				Drug $i$ & Half-life $t_{1/2,i}$ & Value ($\mathrm{h}^{-1}$) & \quad &  Rate constant $\delta_i$   & Value ($\mathrm{min}^{-1}$)  & \quad & Reference\\
				\midrule \midrule 
				1 & $t_{1/2,1}$  &  $\sim 37$   &&  $\delta_1$   & $3.10\times 10^{-4}$ && Sato et al.\cite{sato2006temporal} \\
				2 & $t_{1/2,2}$  & $\sim 6$  &&  $\delta_2$  & $1.93\times 10^{-3}$ && Baselga et al.\cite{baselga2017buparlisib} \\
				3 & $t_{1/2,3}$  & $\sim 2$   && $\delta_3$   & $5.78 \times 10^{-3}$ && Milkiewicz et al.\cite{milkiewicz2011improvement} \\
				4 & $t_{1/2,4}$  & $\sim 1$  && $\delta_4$   & $1.15 \times 10^{-2}$ && Engers et al.\cite{engers2013synthesis} \\
				5 & $t_{1/2,5}$  & $\sim 5$   &&  $\delta_5$  & $2.31 \times 10^{-3}$ && Cameron et al.\cite{cameron2016discovery} \\
				6 & $t_{1/2,6}$  & $\sim 10$&   &   $\delta_6$   & $1.16 \times 10^{-3}$ && Juric et al.\cite{juric2017first} \\
				\midrule
				\bottomrule[1pt]
			\end{tabular}
		}
		\label{tab:param2}
	\end{table}

	\clearpage
	\section*{Supplementary Figures}

\renewcommand{\figurename}{Supplementary Fig.}
\renewcommand{\thefigure}{S\arabic{figure}}

\clearpage

\begin{figure}[tbhp!]
	\centering
	\includegraphics[width=\textwidth]{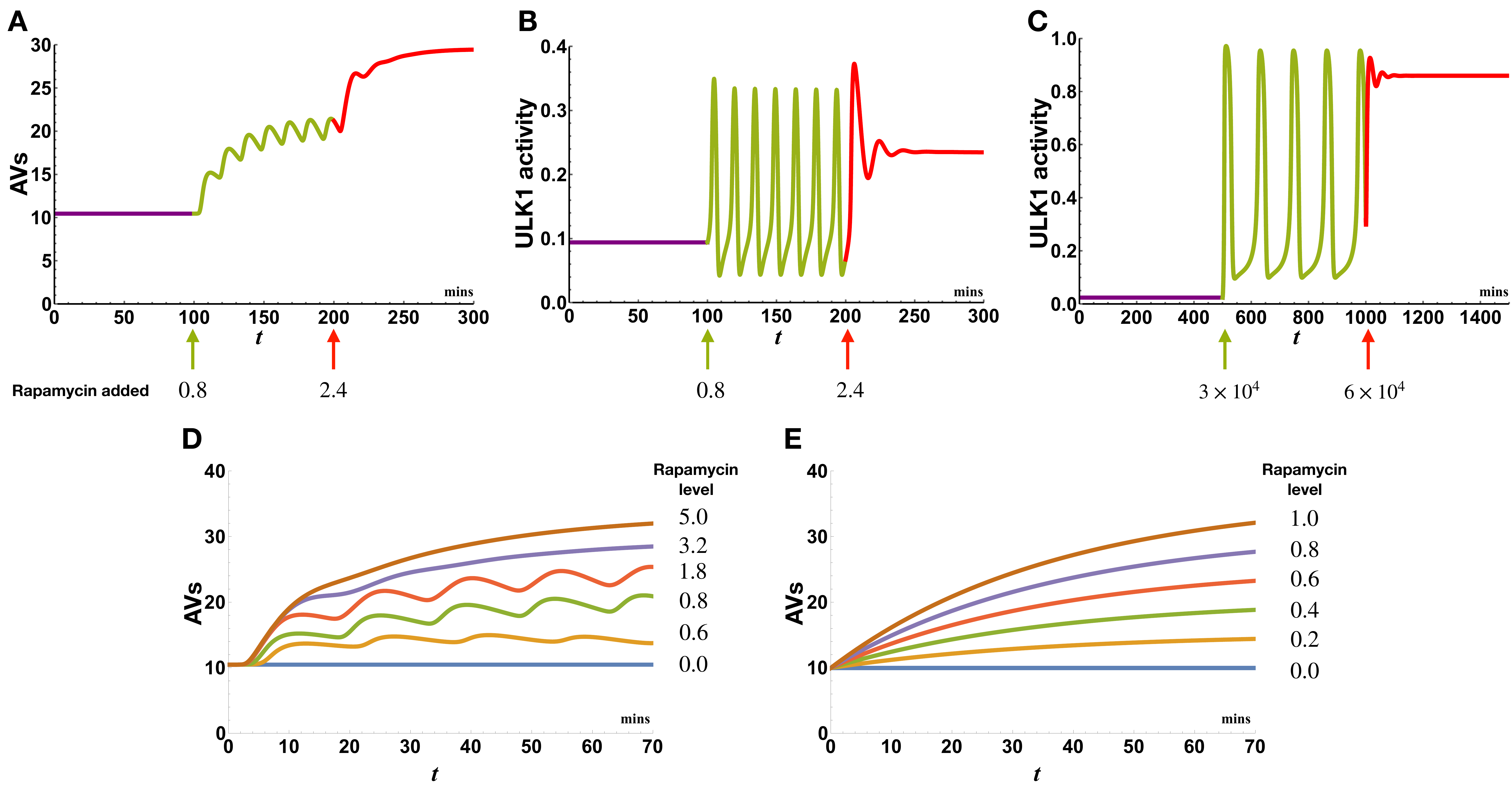}
	\caption{Comparison of simulations based on Eq.\ (1) and simulations based on models of Szyma{\'n}ska et al.\cite{szymanska2015computational} (Ref. 33 in the main text) and Martin et al.\cite{martin2013computational} (Ref. 34 in the main text). (\emph{A}) AV dynamics, $x_5(t)$, predicted by Eq.\ (1). The value of $x_5$ is initially steady and low; the system is perturbed by two additions of rapamycin at time $t=100$ and 200 min, as indicated. (\emph{B}) Dynamics of ULK1 activity, $x_2(t)$, predicted by Eq.\ (1). The conditions considered are the same as those in panel \emph{A}. (\emph{C}) Dynamics of ULK1 activity predicted by the model of Szyma{\'n}ska et al.\cite{szymanska2015computational}. The conditions considered here correspond qualitatively to those considered in panels \emph{A} and \emph{B}. Initially, there is no rapamycin. Later, a low dose of rapamycin is added. Still later, a high dose of rapamycin is added. Note that the models of Eq.\ (1) and Szyma{\'n}ska et al.\cite{szymanska2015computational} have different timescales. This situation is partly a consequence of requiring Eq.\ (1) to reproduce the AV dynamics measured by Martin et al.\cite{martin2013computational}. Szyma{\'n}ska et al.\cite{szymanska2015computational} showed that the qualitative pattern of behavior illustrated here is a robust feature of known regulatory interactions among AMPK, MTORC1, and ULK1 (i.e., the pattern of behavior is insensitive to parameter variations). Furthermore, it should be noted that the model of Szyma{\'n}ska et al.\cite{szymanska2015computational} does not track AVs. Thus, there is no direct comparison to be made with the time course shown in panel \emph{A}. (\emph{D}) AV dynamics predicted by Eq.\ (1). AV production is stimulated by the addition of rapamycin at the (dimensionless) doses indicated in the legend. (\emph{E}) AV dynamics predicted by the model of Martin et al.\cite{martin2013computational}. As in panel \emph{D}, autophagy is induced by the addition of rapamycin at different doses, as indicated in the legend. For further discussion, see ``Formulation of the Model'' in Supplementary Methods.}
	\label{fig:model_vs_model_comparison}
\end{figure}

\clearpage

\begin{figure}[tbhp!]
	\centering
	\includegraphics[width=10cm]{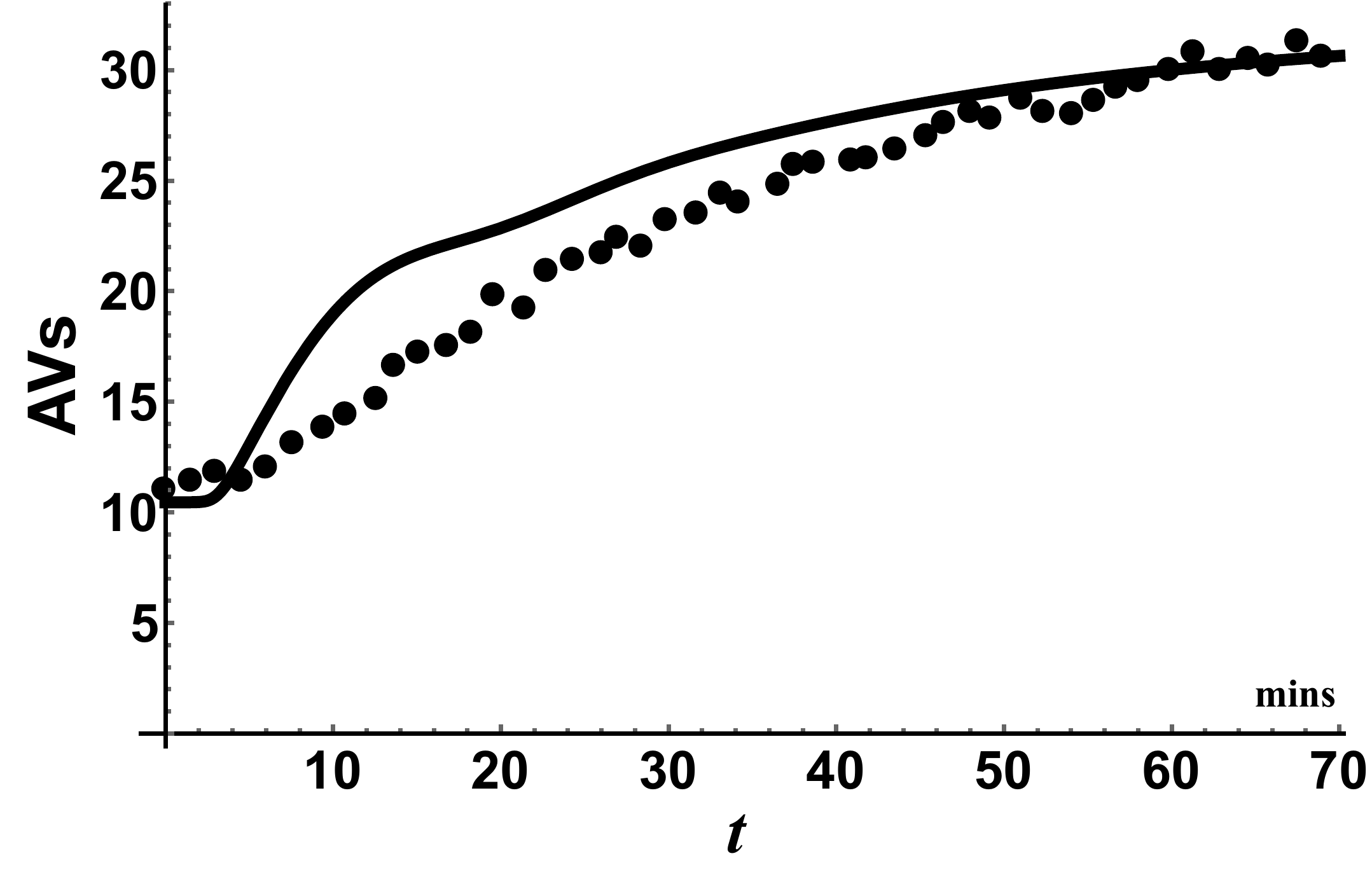}
	\caption{Comparison of simulations based on Eq.\ (1) and data generated by Martin \textit{et al}.\cite{martin2013computational} (Ref. 34 in the main text). We parameterized the model of Eq.\ (1) to roughly reproduce autophagic vesicle (AV) population dynamics reported by Martin et al.\cite{martin2013computational}. Our goal was not to reproduce the observed dynamics exactly but rather to select parameters that yield induction dynamics on a comparable timescale and a comparable maximal range of regulation. The measured dynamics were induced by inhibition of MTORC1 using AZD8055, a catalytic MTOR inhibitor. Dynamics were similar when autophagy was induced using rapamycin\cite{martin2013computational}. The curve corresponds to a simulation based on Eq.\ (1). Each dot corresponds to the average of AV counts measured in a series of fluorescence microscopy experiments\cite{martin2013computational}. The data shown here are taken from Figure 6B in Martin et al.\cite{martin2013computational}. For further discussion, see ``Formulation of the Model'' in Supplementary Methods.}
	\label{fig:model_vs_data_comparison}
\end{figure}
	
	\clearpage
	
	\begin{figure}[tbhp]
		\centering
		\includegraphics[width = 8.7 cm]{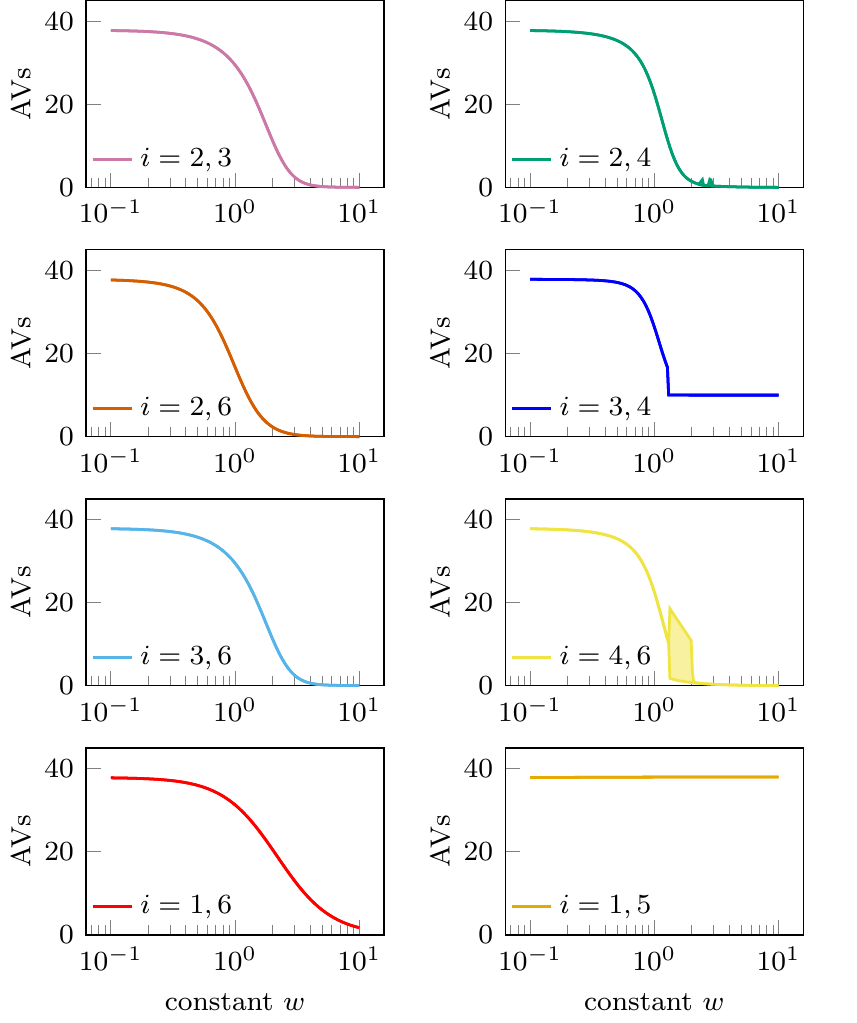}
		\caption{
			The \dual long-time response of the system in the case of time-constant drug concentration perturbations for the parameters $C_ {\text{Nu}} = C_ {\text{En}} = 0.1$.
			Note that when $w$ is small, the system is oscillatory (represented by the shaded region in the panels).
			For each pair of drug, there is some value of $w$ required to overcome the natural oscillatory behavior of the system.
		}
		\label{fig:constinput_2drugs_0101}
	\end{figure}
	
	\clearpage
	
	\begin{figure}[tbhp]
		\centering
		\includegraphics[width = 8.7cm]{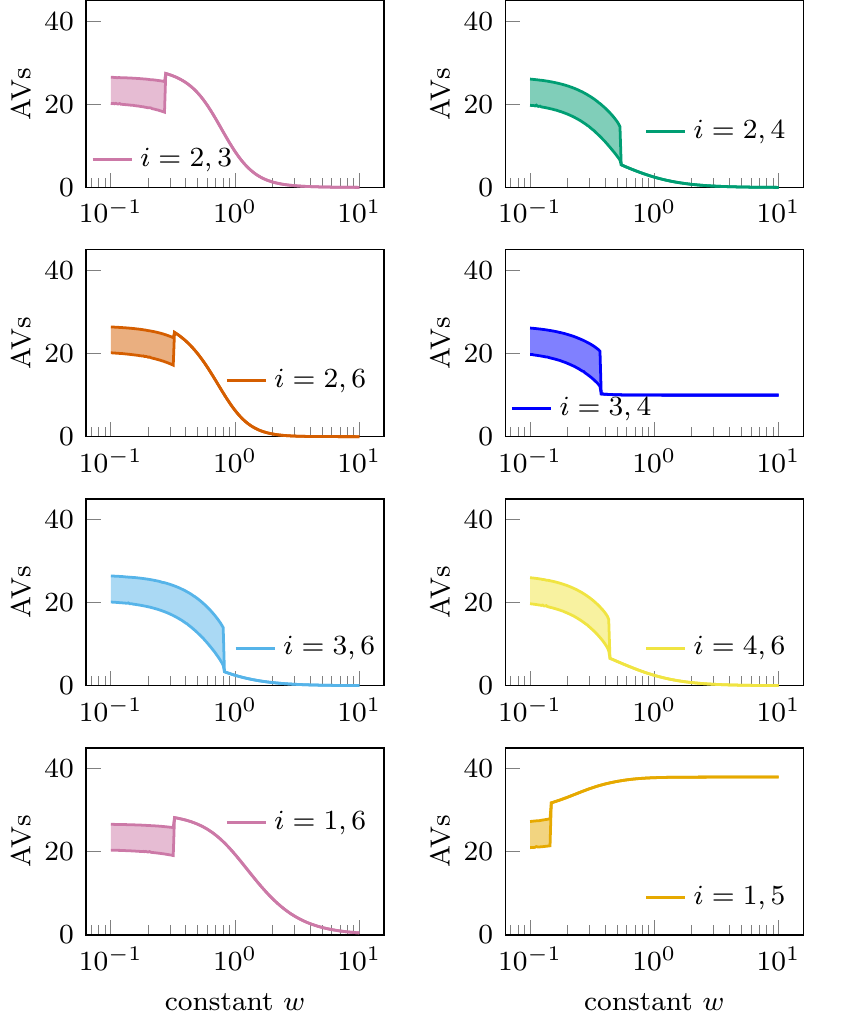}
		\caption{
			The \dual long-time response of the system in the case of time-constant drug concentration perturbations for the parameters $C_ {\text{Nu}} = C_ {\text{En}} = 0.6$.	
		}
		\label{fig:constinput_2drugs_0606}
	\end{figure}
	
	\clearpage
	
	\begin{figure}[t]
		\centering
		\includegraphics[width=17.8cm]{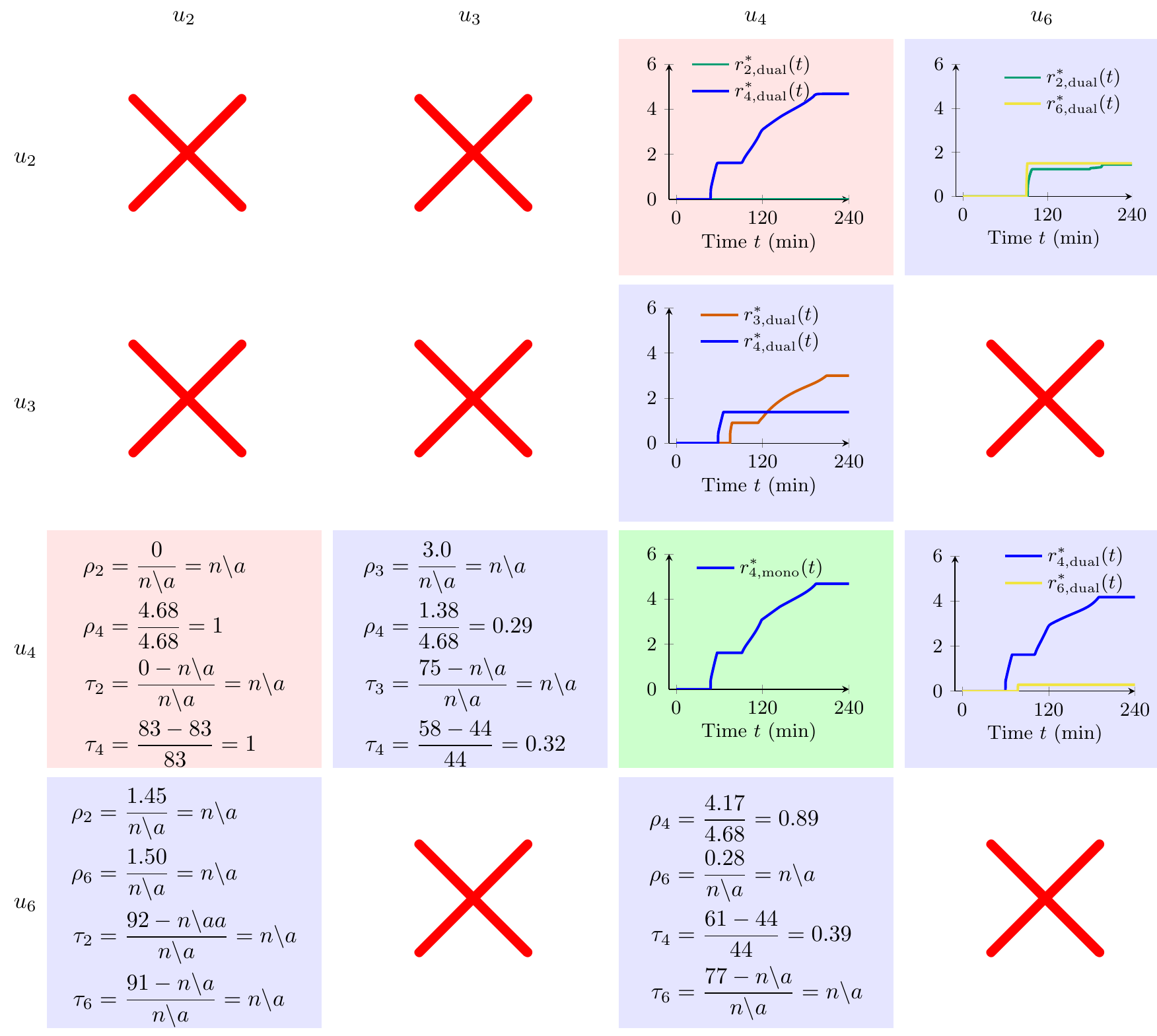}
		\caption{
			The parameter set $C_{\text{Nu}} = C_{\text{En}}=0.1 $.
			The target level of AVs is set $x_5^f = 10$ and the maximum drug concentration is set $w_i^{\max} = 2$.
			The diagonal panels represent \monos while off-diagonal panels represent \duals.
			Super-diagonal panels plot the total drug administered and sub-diagonal panels show the efficiency ratios described in the text of the \duals.
			Those diagonal panels with a red cross correspond to those \monos which are not viable.
			The only viable \mono is $\{4\}$, which is shown with a green background.
			The off-diagonal panel with a red background for \dual $\{2,4\}$ is viable, but it is not efficient as drug $2$ is not activated.
			The other three viable \duals, $\{2,6\}$, $\{3,4\}$, and $\{4,6\}$ are both viable and efficient, shown with a blue background.
		}
		\label{fig:Nu01En01}
	\end{figure}
	
	\clearpage
	
	\begin{figure}[t]
		\centering
		\includegraphics[width=17.8cm]{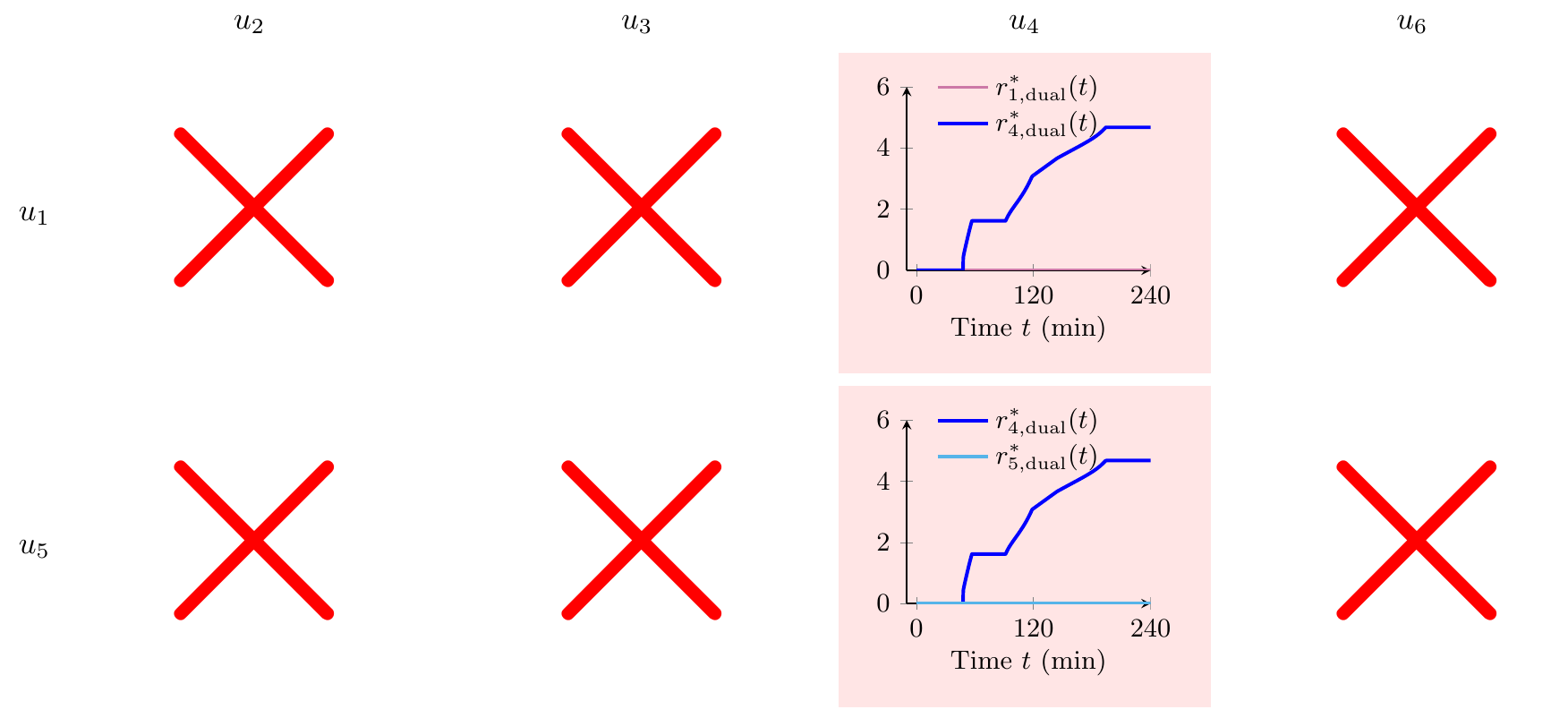}
		\caption{
			The parameter set $C_{\text{Nu}} = C_{\text{En}}=0.1 $.
			The target level of the AVs is set to $x_5^f = 10$ and the maximum drug concentration is set to $w_i^{\max} = 2$.
			Here we consider those \duals which combine one \down drug ($2$, $3$, $4$, or $6$) with one of the \up drugs ($1$ or $5$).
			Most of the \duals are not viable, which is represented with a red cross.
			The two viable \duals, $\{1,4\}$ and $\{4,5\}$, are not viable and so they are shown with a red background.
		}
		\label{fig:Nu01En01_other}
	\end{figure}
	
	\clearpage
	
	\begin{figure}[t]
		\centering
		\includegraphics[width=17.8cm]{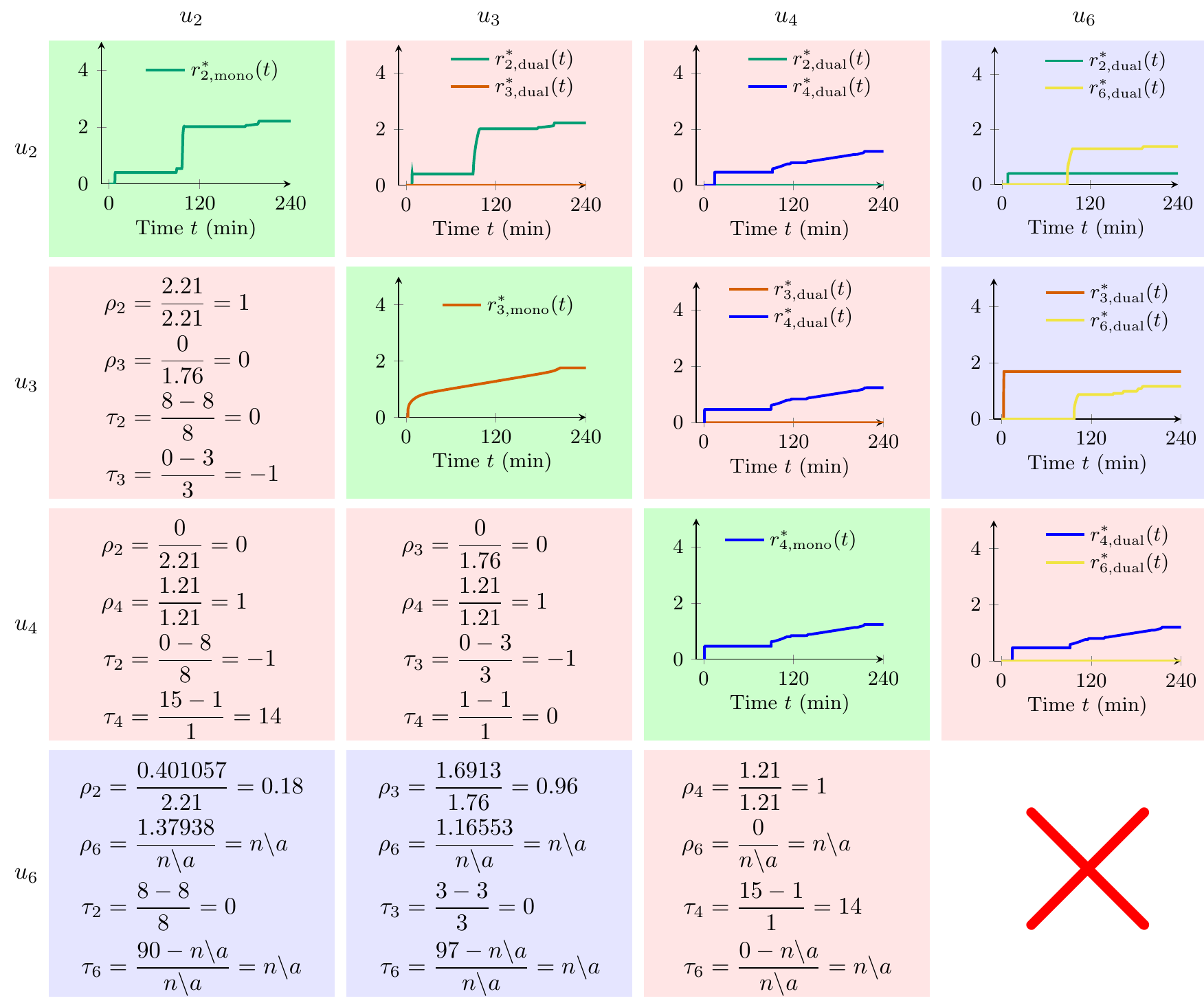}
		\caption{
			The parameter set $C_{\text{Nu}} = C_{\text{En}}=0.6$.
			The target level of the AVs is set to $x_5^f = 10$ and the maximum drug concentration is set to $w_i^{\max} = 2$.
			The diagonal panels $(u_i,u_i)$ (with a green background) show the total drug administered for \monos.
			The red cross on the diagonal panel corresponding to \mono $\{6\}$ represents the fact $\{6\}$ is not viable.
			The upper triangular panels $(u_i,u_j)$, $i < j$, show the total drugs administered for \duals.
			In the lower triangular panels $(u_j,u_i)$, $i < j$, we compare the \duals to their component \monos with the efficiency parameters $\tau$ and $\rho$.
			A red background in an off-diagonal panel represents those \duals which are viable but not efficient with respect to its component \monos.
			A blue background represents those \duals which are both viable and efficient.
		}
		\label{fig:Nu06En06_down}
	\end{figure}
	
	\clearpage
	
	\begin{figure}[tbhp!]
		\centering
		\includegraphics[scale = 1]{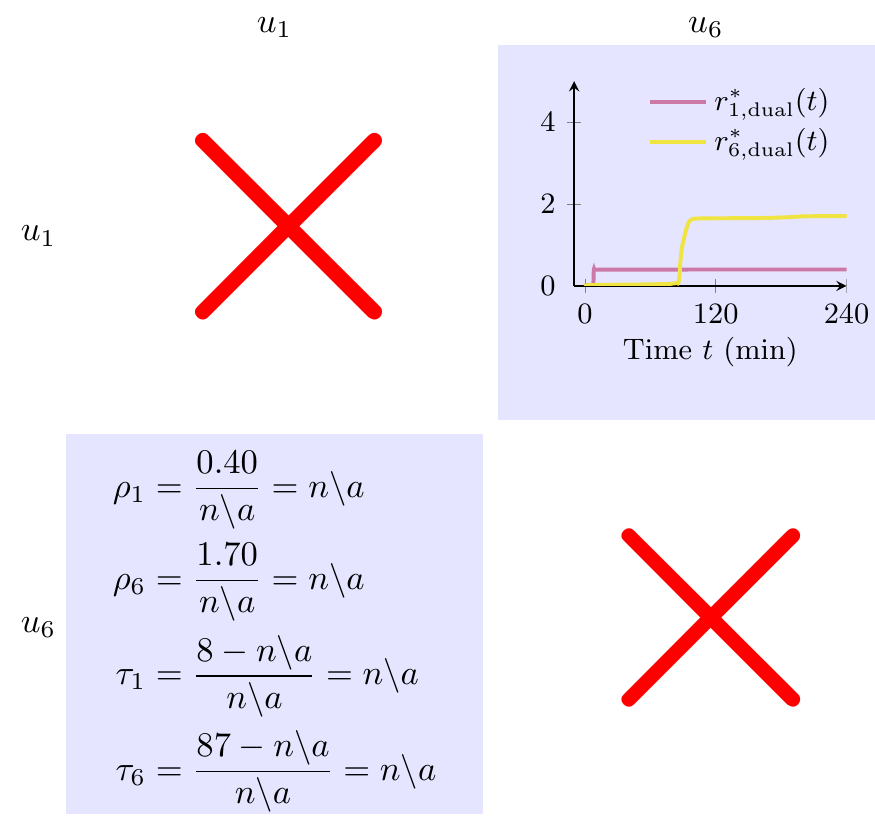}
		\caption{
			The parameter set $C_{\text{Nu}} = C_{\text{En}}=0.6 $.
			The target level of the AVs is set to $x_5^f = 10$ and the maximum drug concentration is set to $w_i^{\max} = 2$.
			The red crosses on the diagonal panels represents the fact that the \monos $\{1\}$ and $\{6\}$ are not viable.
			On the other hand, the \dual $\{1,6\}$ is both viable and efficient.
			The viable \duals composed of two \monos which are not viable alone are the type of \duals we find most interesting as they are not obvious when analyzing the \monos alone.
			In the lower triangular panel we compare the \dual to its component \monos with respect to the efficiency ratios $\rho$ and $\tau$.
		}
		\label{fig:Nu06En06_down_2}
	\end{figure}
	
	\clearpage
	
	\begin{figure}[tbhp!]
		\centering
		\includegraphics[scale = 1]{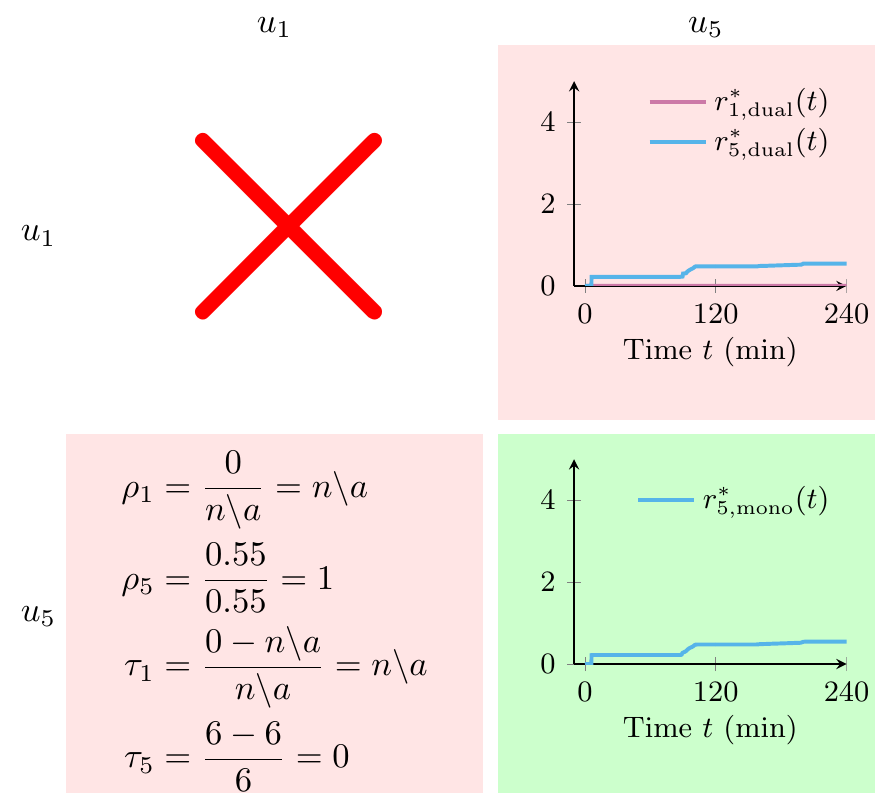}
		\caption{
			The parameter set $C_{\text{Nu}} = C_{\text{En}}=0.6 $.
			The target level of the AVs is set to $x_5^f = 10$ and the maximum drug concentration is set to $w_i^{\max} = 2$.
			The diagonal panels represent the \monos $\{1\}$ and $\{5\}$.
			A red cross on the diagonal panel for \mono $\{1\}$ represents the fact $\{1\}$ is not viable.
			On the other hand, \mono $\{5\}$ is viable (shown with a green background).
			The \dual $\{1,5\}$ is viable (total drug administered is shown with the red background in the upper triangular panel) but is not efficient.
			The inefficiency is shown in the lower triangular panel with the efficiency ratios $\rho_5 = 1$.
		}
		\label{fig:Nu06En06_up}
	\end{figure}
	
	\begin{figure}[t]
		\centering
		\includegraphics[width=17.8cm]{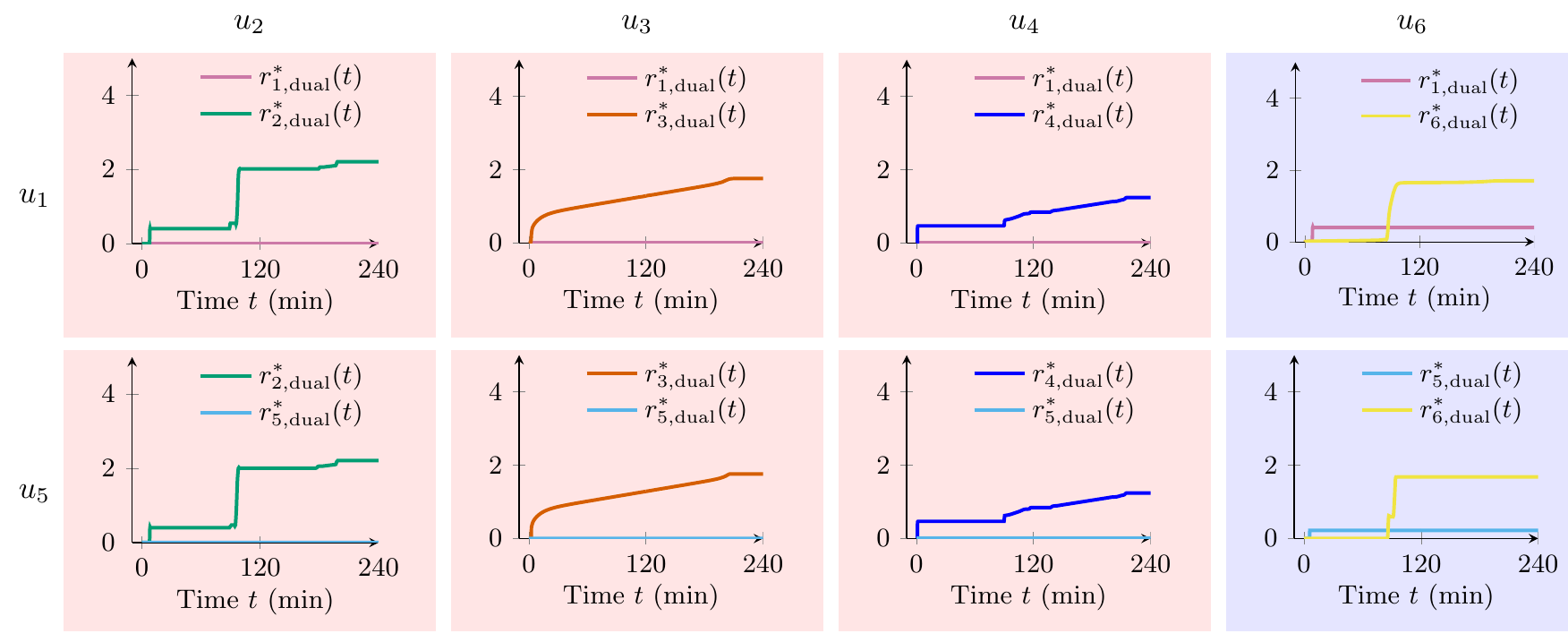}
		\caption{
			The parameter set $C_{\text{Nu}} = C_{\text{En}}=0.6$.
			The target level of AVs is set to $x_5^f = 10$ and the maximum drug concentration is set to $w_i^{\max} = 2$.
			Here we consider those \duals compose of one \down drug ($2$, $3$, $4$, or $6$), and one \up drug ($1$ or $5$).
			Those panels with a red background represent \duals which are viable but not efficient while the two \duals $\{1,6\}$ and $\{5,6\}$ are efficient.
			In fact, as seen before, neither the component \mono $\{6\}$ nor the \up drugs are viable for this parameter set, so these efficient \duals are particularly interesting as they could not be found when analyzing the \monos alone.
		}
		\label{fig:Nu06En06_other}
	\end{figure}
	
	\clearpage
	
	\begin{figure}[tbhp!]
		\centering
		\includegraphics[width = 8.7cm]{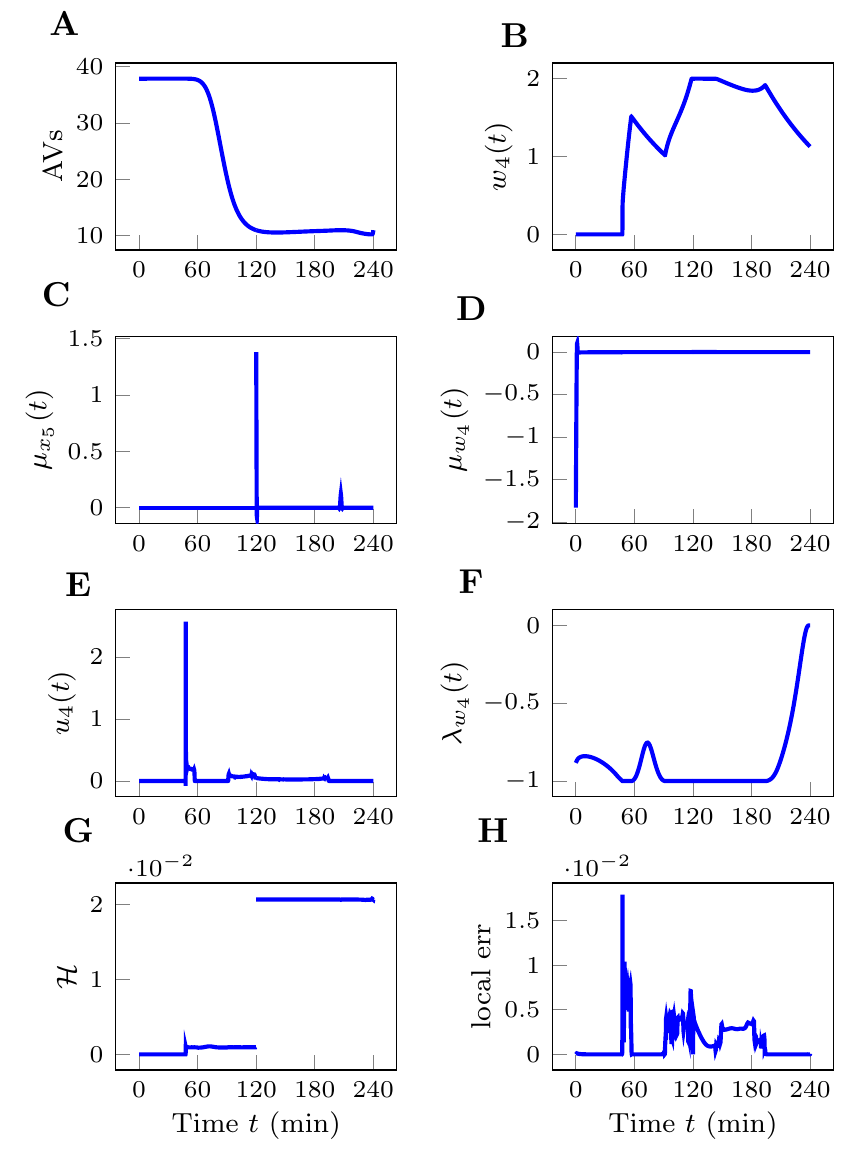}
		\caption{ 
			a) The optimal time evolution of the amount of AVs. 
			b) The optimal time evolution of the drug concentration $w_4(t)$.
			c) The time evolution of the path covector $\mu_{x_5}$ associated with the upper bound applied to $x_5(t)$.
			d) The time evolution of the path covector $\mu_{w_4}$ associated with the state $w_4(t)$.
			e) The optimal time evolution of the drug $u_4(t)$.
			f) The costate $\lambda_{w_4}(t)$ associated with the state  $w_4(t)$.
			g) The time evolution of the lower Hamiltonian $\mathcal{H}$.
			h) The relative local discretization error at each time $t$.}
		\label{fig:VV}
	\end{figure}

	\bibliographystyle{naturemag-doi}